\documentclass[a4paper,11pt]{article}
\pdfoutput=1

\usepackage[normalem]{ulem}
\usepackage{jcappub} 
\usepackage{amsmath}
\usepackage{amssymb}
\usepackage{color}
\usepackage{graphicx}
\usepackage{comment}
\usepackage{multirow}
\usepackage[dvipsnames]{xcolor}

\usepackage[utf8]{inputenc}

\usepackage{caption}

\include{journals}

\graphicspath{{images/}}

\newcommand{\lam}{$\Lambda$}

\newcommand{\ol}{$\Omega_\Lambda$}
\newcommand{\om}{$\Omega_\mathrm{M}$}
\newcommand{\ost}{$\Omega_*$}
\newcommand{\ok}{$\Omega_K$}
\newcommand{\wo}{$w_0$}
\newcommand{\phiroot}{\Omega_\phi^{1/2}}
\newcommand{\ome}{$\Omega_e$}
\newcommand{\onu}{$\Omega_\nu$}
\newcommand{\ophi}{$\Omega_\phi$}
\newcommand{\rcmbao}{$f_{\mathrm{CMB/BAO}}$}
\newcommand{\ods}{$\Omega_{\mathrm{ds}}$}
\newcommand{\ho}{$H_0$}

\newcommand{\dlz}{$\Delta \mathrm{ln} \, Z$}

\newcommand{\hz}{$H(z)$}

\newcommand{\wz}{$w(z)$}
\newcommand{\la}{$l_A$}

\newcommand{\rsd}{$r_s(z_d)$}
\newcommand{\rsst}{$r_s(z_*)$}
\newcommand{\dw}{$\delta w_0$}
\newcommand{\pl}{\emph{Planck}}

\def\gsim{\mathrel{\rlap{\lower 4pt \hbox{\hskip 1pt $\sim$}}\raise 1pt \hbox {$>$}}}
\def\lsim{\mathrel{\rlap{\lower 4pt \hbox{\hskip 1pt $\sim$}}\raise 1pt \hbox {$<$}}}

\begin{document}

\title{\boldmath Narrowing down the possible explanations of cosmic acceleration with geometric probes}

\author[a]{Suhail Dhawan,}
\affiliation[a]{Oskar Klein Centre, Department of Physics, Stockholm University, SE 106 91 Stockholm, Sweden}
\emailAdd{suhail.dhawan@fysik.su.se}

\author[a]{Ariel Goobar,}
\emailAdd{ariel@fysik.su.se}

\author[a]{Edvard M{\"o}rtsell,}
\emailAdd{edvard@fysik.su.se}

\author[a]{Rahman Amanullah,}
\emailAdd{rahman@fysik.su.se}

\author[a]{and Ulrich Feindt}
\emailAdd{ulrich.feindt@fysik.su.se}


\abstract{
Recent re-calibration of the Type Ia supernova (SNe~Ia) magnitude-redshift relation combined with cosmic microwave background (CMB) and baryon acoustic oscillation (BAO) data have provided excellent constraints on the standard cosmological model. Here, we examine particular classes of alternative cosmologies, motivated by various physical mechanisms, e.g. scalar fields, modified gravity and phase transitions to test their consistency with observations of SNe~Ia and the ratio of the angular diameter distances from the CMB and BAO. Using a model selection criterion for a relative comparison of the models (the Bayes Factor), we find moderate to strong evidence that the data prefer flat \lam CDM over models invoking a thawing behaviour of the quintessence scalar field. However, some exotic models like the growing neutrino mass cosmology and vacuum metamorphosis still present acceptable evidence values. The bimetric gravity model with only the linear interaction term can be ruled out by the combination of SNe~Ia and CMB/BAO datasets whereas the model with linear and quadratic interaction terms has a comparable evidence value to standard \lam CDM. Thawing models are found to have significantly poorer evidence compared to flat \lam CDM cosmology 
under the assumption that the CMB compressed likelihood provides an adequate description for these non-standard cosmologies.  
We also present estimates for constraints from future data and find that geometric probes from oncoming surveys can put severe limits on  non-standard cosmological models.
}

\maketitle

\section{Introduction}
Close to two decades after the discovery of the accelerated expansion of the universe \cite{1998AJ....116.1009R,1999ApJ...517..565P}, little is known about the underlying dark energy physics responsible for the observed phenomenon. Improved measurements of Type Ia supernovae \cite[hereafter SN~Ia;][]{2014A&A...568A..22B}, the cosmic microwave background \cite[CMB;][]{2016A&A...594A..13P} and baryon acoustic oscillations \cite[BAO;][]{2015PhRvD..92l3516A} show a continued consistency with Einstein's cosmological constant, \lam, a spatial homogeneous energy density with an equation of state (EoS), or pressure to density ratio $w = -1$ \cite{2014A&A...568A..22B,2016A&A...594A..13P}. However, there are two striking problems with the cosmological constant \cite[see][for a detailed discussion]{2000astro.ph..5265W}. Firstly, in context of quantum field theory, there isn't yet an understanding for why the observed density of the vacuum, $\rho_{vac} = \frac{\Lambda}{8\pi G} \sim (0.003 \mathrm{eV})^4$\ol\, is not of the order of M$_{\mathrm{pl}}^{4}$ or at least the supersymmetry break scaling M$_{\mathrm{SUSY}}^{4} \sim 1$ TeV$^4$, both of which yield a discrepancy of several orders of magnitude.  
Secondly, the present day energy density of matter (\om) and the cosmological constant (\ol) are of the same order of magnitude, despite having extremely different dependence on the scale factor, which known as the \emph{cosmological coincidence problem} \cite[in][they are referred to as the `old' and `new' cosmological constant problems respectively]{2000astro.ph..5265W}. 

Dark energy models beyond the cosmological constant have been invoked to solve either one or both of these problems. Such models are constructed from different physical mechanisms including, but not limited to, a light scalar field $\phi$ with an associated potential $V(\phi)$,  modifications to the equations of general relativity \cite[see][and references therein]{2006IJMPD..15.1753C,2013CQGra..30u4003T}, braneworld gravity, mass-varying neutrinos, phantom energy etc. \cite[see][for a review of the possible physical causes of cosmic acceleration]{2009PhRvD..79h3517C}. Some of these  physical mechanisms can be invoked to explain the recent possibly tension in \ho\, measurements from the CMB \cite{2016A&A...594A..13P} and local, astrophysical measurements \cite{2016ApJ...826...56R}. 
In this study we analyse parameter constraints from models with various physical motivations, some of which were presented in \cite{2009ApJ...695..391R}  (along with recent parametrisations for dark energy evolution e.g. \cite{2011MNRAS.416..907G}). We analyse recent SN~Ia, CMB and BAO data by using an almost completely model independent approach to obtain parameter constraints for each model and by presenting a full Bayesian model selection to compare the different models. 

To test the different dark energy models, we combine constraints from complementary datasets, namely, the SN~Ia luminosity distance-redshift relation, the CMB compressed likelihood and the angular scale in the local universe given by BAO observations. 
Previously, there have been studies to explore non-standard cosmologies \cite[different from the models tested here; e.g.][]{2007ApJ...666..716D,2009ApJ...703.1374S}, wherein the different models were compared using the Bayesian Information Criterion \cite[BIC;][]{1978AnSta...6..461S}. Due to the nature of information criteria used, both studies find a preference for models with fewer parameters and conclude that the best fit cosmology is a flat \lam CDM \cite[although][find that flat \lam CDM is not the most favoured cosmology if the SN data are analysed using a different light curve fitting technique]{2009ApJ...703.1374S}. In \cite{2009ApJ...695..391R} several different dark energy models were compared using a $\chi^2$ statistic and some models were included, while others survive by approaching the cosmological constant regime and a few are also better fit to the data than  \lam CDM. In \cite{2012SCPMA..55.2244Z}\,  a series of modified gravity models are tested. Using  $\chi^2$ and BIC values for comparing the different models, the authors find find that, although the SNe by themselves prefer a Dvali-Gabadadze-Porratti cosmology \cite[DGP;][]{2000PhLB..485..208D}, a combination with complementary data (e.g. CMB, BAO) shows  \lam CDM to be the most favoured model. Models motivated by modified gravity and with different dark energy EoS parametrisations are tested in \cite{2008ApJ...675....1K}, wherein the authors conclude that there is a preference for flat \lam CDM. 

Here, we examine a series of cosmological models with different physical motivations using the latest SN~Ia, CMB and BAO datasets in order to test whether the standard ``concordance'' cosmology is indeed the most favourable fit to the data. We present observational constraints on the parameters and analyse the regions of the parameter space allowed and excluded by the data. We employ a model selection criterion, namely, the Bayes Factor, which is the ratio of the Bayesian evidence \cite{2004AIPC..735..395S}.
In Section~\ref{sec:data} we describe the datasets used in our work. In section~\ref{sec:param_est} we constrain the parameter space for the models and compare them in~\ref{sec:mod_sel} and discuss forecasts for future surveys in Section~\ref{sec:forecast}. Finally, we discuss our results in the section~\ref{sec:disc}.

\section{Data}
\label{sec:data}

\subsection{Type Ia supernovae}
\label{ssec:sn}
In this work, we use the SNe~Ia luminosity distance relation from the ``Joint Lightcurve Analysis" \cite[JLA][]{2014A&A...568A..22B} containing 740 spectroscopically confirmed SNe~Ia
Theoretically, the distance modulus predicted by the homogeneous and isotropic, flat Friedman-Robertson-Walker (FRW) universe is given by

\begin{equation}
\mu(z; \theta) = 5 \mathrm{log_{10}} \left( \frac{d_L}{10 \mathrm{Mpc}} \right) + 25
\label{eq:mu_sne}
\end{equation}
where $z$ is the redshift, $\theta$ are the cosmological parameters and $d_L$ is given by 

\begin{equation}
d_L = \frac{c(1+z)}{H_0 \sqrt{|\Omega_k|}}\, \mathrm {sinn}\, \left( \sqrt{|\Omega_k|} \int^{z}_{0} \frac{dz^{'}}{E(z^{'})} \right)
\label{eq:lum_dist}
\end{equation}
where sinn (x) = \{ sin(X), x, sinh(x) \} for close, flat and open universes, respectively. $E(z) = H(z)/H_0$ is the normalised Hubble parameter for each cosmological model.

The Hubble parameter has a generic form of 
\begin{equation}
\frac{H(z)^2}{H_0^2} = \Omega_M (1+z)^3 + \Omega_{DE}(z) + \Omega_K(1+z)^2, 
\label{eq:norm_hubbleparameter}
\end{equation}
where 

\begin{equation}
\mathrm{\Omega_{DE}(z)} = \Omega_\Lambda \mathrm{exp} \left[ 3 \int_0^z \frac{(1+w(x) ) }{1 + x} dx \right],
\label{eq:ode_z}
\end{equation}
where $w(z)$ is the dark energy EoS. For the flat and non-flat \lam CDM cases tested here, $w$ is a constant set to -1 (corresponding to a cosmological constant), whereas for the more non-standard cosmology we present the \hz\, or \wz\, expressions in the respective sections. 

Observationally, the distance modulus is calculated from the SN~Ia peak apparent magnitude ($m_B$), light curve width ($x_1$) and colour ($c$)

\begin{equation}
\mu_{obs} = m_B - (M_B - \alpha x_1 + \beta c),
\label{eq:obs_distmod}
\end{equation}
where $M_B$ is the absolute magnitude of the SN~Ia. Following \cite{2014A&A...568A..22B}, we apply a step correction ($\Delta_M$) for the host galaxy stellar mass ($M_{\mathrm{stellar}}$). We note that $\alpha$, $\beta$, $M_B$ and $\Delta_M$ are nuisance parameters in the fit for the cosmology. 

The expression for $\chi^2$ is given by 
\begin{equation}
\chi_{\mathrm{SN}}^2 = \Delta^T C_{\mathrm{SN}}^{-1} \Delta, 
\end{equation}
where $\Delta = \mu - \mu_{obs}$ and C is the complete covariance matrix described in \cite{2014A&A...568A..22B}.

\subsection{Cosmic Microwave Background}
\label{ssec:cmb}
In this section we describe the CMB data used to complement the SNe~Ia data discussed above. \cite{2002PhRvD..66f3007K} and \cite{2007PhRvD..76j3533W} discuss the information from the CMB power spectrum can be compressed into a few parameters. Here, we use the compressed likelihood from \pl\, 2015 \cite{2016A&A...594A..13P} which includes the CMB shift parameter, R, \cite{1997MNRAS.291L..33B,1999MNRAS.304...75E} and the position of the first peak in the CMB power spectrum,  \la\, \cite{2006ApJ...650....1W,2007A&A...471...65E,2009ApJS..180..330K}, expressed as 
\begin{equation}
R = \sqrt{\Omega_M H_0^2} d_A(z_*)/c,
\label{eq:cmb_shift}
\end{equation}
\begin{equation}
l_A = \pi \frac{d_A(z_*)}{r_s(z_*)}, 
\label{eq:cmb_la}
\end{equation}
where \om\, is the dimensionless matter density, $d_A(z_*)$ is the comoving angular diameter distance at the redshift of decoupling, $z_*$. The expression for redshift of decoupling is given by the fitting formula of \cite{1996ApJ...471..542H} 

\begin{equation}
z_* = 1048 \cdot [1 + 0.00124(\Omega_B h^2)^{-0.738}][1+g_1 (\Omega_M h^2)^{g_2}],
\label{eq:decoup_redshift}
\end{equation}
and 
\begin{equation}
g_1 = \frac{0.0783(\Omega_B h^2)^{-0.238}}{1+39.5(\Omega_B h^2)^{0.763}},
\label{eq:g1_shiftcoef}
\end{equation}
\begin{equation}
g_2 = \frac{0.560}{1+21.1(\Omega_B h^2)^{1.81}}, 
\label{eq:g2_shiftcoef}
\end{equation}
where $\Omega_B$ is the dimensionless baryon density and $h = H_0/100$. For our calculations we set the $\Omega_B$ from \pl\, \cite{2016A&A...594A..13P} to 0.0448 (which, along with \om\, sets the value for the recombination redshift, $z^*$.). 

From the \pl\, 2015 data (marginalised over the lensing amplitude, A$_L$), we use $R$ = 1.7382 $\pm$ 0.0088 and \la\, = 301.63 $\pm$ 0.15 respectively. The covariance between them is 0.64 \cite[see Table 4 of reference][]{2016A&A...594A..13P} 
Using these two parameters in combination can reproduce the fit from the full CMB power spectrum \cite[see][for caveats]{2007A&A...471...65E}. We note, however, that although the CMB shift, $R$, alone has commonly been used for cosmological constraints, it may not be appropriate for testing alternate cosmologies \cite[see][]{2007A&A...471...65E}. This is because parameters close to the $w$CDM model have been used to derive these constraints from the full power spectrum \cite[see section 5.1.6 of reference][]{2016A&A...594A..14P}.

From the discussion in \cite{2016A&A...594A..14P}, it is clear that the strong model dependence of the CMB shift parameter makes it unfeasible for complementing the SN~Ia constraints on non-standard cosmology. The authors also state that the compressed CMB likelihood (a combination of $R$ and \la) is feasible for smooth dark energy models but the parameters, especially the shift parameter, $R$, are sensitive to growth perturbations. This is most clearly seen in the ``dark degeneracy'' \cite{2009PhRvD..80l3001K}, i.e. the possibility to absorb part of the dark matter into the dark energy, which changes \om\, without affecting observables. Hence, the compressed likelihood is not applicable for models that involve a modification of gravity. 
The models tested here include diverse physical motivations, including modifications to gravity. We therefore do not use the CMB compressed likelihood for all models, only a subset where the compression is applicable. 

Thus, for a complete model comparison of all the different scenarios detailed here, we use the position of the first CMB acoustic peak, \la, combined with the BAO scale to complement the SN~Ia data. A description of this parameter is presented in Section~\ref{ssec:fratio}, after the description of the BAO data in Section~\ref{ssec:bao}.  

\subsection{Baryon Acoustic Oscillation}
\label{ssec:bao}

The detection of the characteristic scale of the baryon acoustic oscillations (BAO) in the correlation function of different matter distribution tracers provides a powerful standard ruler to probe the angular-diameter-distance versus redshift relation. BAO analyses usually perform a spherical average  constraining a combination of the angular scale and redshift separation

\begin{equation}
d_z = \frac{r_s(z_{\mathrm{drag}})}{D_V(z)}, 
\label{eq:dzbao}
\end{equation}
with 
\begin{equation}
D_V(z) = \left[(1+z)^2 D_A(z)^2 \frac{cz}{H(z)} \right] ^{1/3}, 
\label{eq:dvz}
\end{equation}
where $D_A$ is the angular diameter distance and \hz\, is the Hubble parameter. 
$r_s(z_{drag})$, is calculated using the fitting formula from \cite{1998ApJ...496..605E}. 
For our analyses, we follow the method of \cite{2016A&A...594A..13P} and use four measurements from 6dFGS at $z_{\mathrm{eff}} = 0.106$, the recent SDSS main galaxy (MGS) at $z_{\mathrm{eff}}$ = 0.15 of \cite{2015MNRAS.449..835R} and $z_{\mathrm{eff}}$ = 0.32 and 0.57 for the Baryon Oscillation Spectroscopic Survey (BOSS) ``LOWZ'' and ``CMASS'' surveys respectively \cite{2014MNRAS.441...24A} with a precision of 3$\%$, 4$\%$, 2$\%$  and 1$\%$ respectively \cite[see also ][for a discussion of the complementary BAO data]{2016EPJC...76..588X}
We consider a BAO prior of the form 
\begin{equation}
\chi^2_{\mathrm{BAO}} = (d_z - d_z^{\mathrm{BAO}})^T C_{\mathrm{BAO}}^{-1} (d_z - d_z^{\mathrm{BAO}}), 
\label{eq:chi_bao}
\end{equation}
with $d_z^{\mathrm{BAO}}$ = [0.336, 0.2239,0.1181, 0.07206] and $C_{\mathrm{BAO}}^{-1}$ = diag(4444.44,    14071.64,   183411.36,
        2005139.41). 
        
We note that the WiggleZ survey also present three distance measurements at \cite[see][]{2014MNRAS.441.3524K,2011MNRAS.416.3017B}. However, since the WiggleZ volume partially overlaps with that of the BOSS CMASS sample, and the correlations have not been quantified, we do not include the WiggleZ results in this study. 


\subsection{CMB/BAO ratio}
\label{ssec:fratio}

As discussed above, the CMB compressed likelihood has been calculated from the power spectrum assuming a cosmological model (namely, wCDM). Hence, although it is applicable for certain kinds of models tested here, it would not be a suitable measurement to use for all models in this study, especially those that involve a modification of gravity or an early time contribution of dark energy, namely the vacuum metamorphosis and growing neutrino mass models. Thus, for a uniform model comparison wherein all the models have been constrained with the same combination of datasets, we aim to use a more model independent approach, similar to \cite{2009ApJ...703.1374S}. We combine the position of the first CMB acoustic peak, $l_A$, with the estimate of $d_z$,  to get ratio of the  comoving distance at the redshift of decoupling and the dilation scale at low redshift. 

Hence, the ratio of the angular diameter distance at last scattering surface and at low $z$ is given by \cite[we use the letter $f$ to denote the ratio following reference ][]{2012ApJ...744..176L}.  

\begin{equation}
f = \frac{d_A(z_*)}{D_V(z)} = \frac{l_A}{\pi d_z} \cdot \frac{r_s(z_d)}{r_s(z_*)} ,
\label{eq:cmbao_ratio}
\end{equation}
where the ratio of the sound horizons at $z_*$ and $z_d$ is calculated from the  \rsst\, and \rsd\, values of \pl\, 2015 using the CMB temperature power spectrum and low-$l$ polarisation \cite{2016A&A...594A..13P}. We note that this ratio was calculated for a background \lam CDM cosmology for the evolution between the two redshifts. We expect this to be a good approximation for the models tested in this study since the redshift difference between the decoupling and the drag epoch is relatively small. Moreover, the sound horizon at decoupling and drag is mostly governed by the fractional difference between the number of photons and baryons.

Combining the four BAO measurements described in Section~\ref{ssec:bao} with $l_A$ from \cite{2016A&A...594A..13P}, the values of the ratio (\rcmbao) at each of the three redshifts is [31.6895,  21.1169,  11.1385,   6.8472].

The $\chi^2$ is

\begin{equation}
\chi_{\mathrm{CMB/BAO}}^2 = \Delta^T C_{\mathrm{CMB/BAO}}^{-1} \Delta
\label{eq:chi_cmbao}
\end{equation} 	
where $\Delta = f^{th}_{\mathrm{CMB/BAO}} - f^{obs}_{\mathrm{CMB/BAO}}$ and the covariance matrix C is diag([1.6384, 0.9441, 0.2988, 0.1149]). We note that the precision of the ratio measurements is reduced compared to either the BAO or CMB measurements, but, as detailed above, this measurement is almost completely model independent \cite[except for the ratio of \rsst\, and \rsd\, which as we discuss above is a small effect][present a discussion of the $f$ ratio]{2009ApJ...703.1374S}.

We note that the measurement of this ratio is completely independent of the absolute expansion scale, i.e. the Hubble constant, \ho. Hence, the addition of the CMB and BAO data in this form doesn't break the existing degeneracy between the absolute magnitude of SNe~Ia, $M_B$, and \ho. 

\smallskip

We emphasize here that the aim of introducing the CMB/BAO ratio is to have a model independent measurement to perform a model selection for the different cosmologies. Where applicable, we also present the constraints from the CMB compressed likelihood and BAO angular scale separately since these constraints are expected to be more precise than the constraints from the CMB/BAO ratio, $f$. Since the datasets correspond to independent measurements, we combine them by adding the $\chi^2$.


\begin{table*}
\centering
\caption{List of Models tested in this study}
\begin{tabular}{|l|l|c|}
\hline\hline
Model & Parameters & Motivation  \\
\hline
(flat) \lam CDM (f \lam) & \om & Gravity, Zeropoint \\ 
$\Lambda$CDM (\lam) & \om, \ok & Gravity, Curvature as a free parameter    \\ 
&& \\
Vacuum Metamorphosis (VM) & \om, \ost & Quantum fluctuations as induced gravity     \\
Doomsday (DD) & \om, \wo & Simple Scalar Field  \\
One parameter slow-roll dark energy (SR) & \om, $\delta w_0$ & Analogous to inflation \\ 
Algebraic Thawing (AT) & \om, \wo, $p$ &  Generic Evolution   \\
PNGB   & \ophi, \wo, $K$ & Technical Naturalness  \\
Growing $\nu$ mass (G$\nu$) & \om, \ome, \onu & Solves Coincidence Problem   \\
Bimetric-Linear (BL) & \om & Acceleration without vacuum energy\\
Bimetric-Linear and Quadratic (BQ) &\om, $B_2$& Acceleration without vacuum energy \\
&& \\
\hline
\end{tabular}
\label{tab:modlist}
\end{table*}

\section{Parameter Estimation}
\label{sec:param_est}
Informative limits on dark energy can be achieved by constraining the free parameters of the theory and the resulting degeneracies in the cosmological model. One cosmological parameter in the analyses is the present day matter density \om\, (the expressions for the various datasets and how they constrain \om\, is presented in Section~\ref{sec:data}). For the flat \lam CDM model this is the only parameter determining the SN magnitude-redshift relation and the CMB/BAO ratio. 


In this section we investigate various two parameter and three parameter models, discussing their physical motivation, their features of interest and observational constraints. \cite{2005PhRvD..72d3509L} argue that even next generation experiments for low- and high-redshift cosmological probes will only be able to constrain three parameter models for the dark energy evolution. Hence, in this study, we restrict the models tested to have two dark energy parameters (and a total of three parameters, including \om), relating to the value of the equation of state at a specific epoch (e.g. at present day, \wo) and a parameter describing its behaviour with time ($w^{'}$ or $w_a$). Since more elaborate dark energy parametrisations can be useful for testing biases, but not for robust measurements of dark energy parameters. 

Models tested in this study are listed in Table~\ref{tab:modlist}. For the first time, in this study we present the parameter estimation and model selection for this set of cosmologies using a model independent approach of combining the SN distance-redshift relation with $f$, the CMB/BAO  angular scale ratio. We note that some models motivated by a modification of Einstein's gravity e.g. DGP \cite{2000PhLB..485..208D}, have ghost modes which make them theoretically  not viable \cite{2006PhRvD..73d4016G}. Standard Chaplygin gas \cite[][and references therein]{2002PhRvD..66d3507B} has been shown as inconsistent with existing data \cite[for e.g., see][]{2003PhRvD..68b3515B,2006PhLB..642..432F,2007ApJ...666..716D,2009ApJ...703.1374S,2009ApJ...695..391R} and thus, have not been inspected here. 

The nuisance parameters for fitting the SN~Ia data with the models in this study are in excellent agreement with the values for standard cosmologies \cite[Tables 13 - 15 in reference][]{2014A&A...568A..22B}. 

Note that all models in Section~\ref{ssec:doom} to ~\ref{ssec:grownu} assume flatness and hence, for model selection are compared to the flat \lam CDM standard cosmology.

\subsection{Vacuum Metamorphosis}
Solutions to explain the accelerated expansion of the universe include the cosmological constant, light scalar fields and modifications to Einstein's gravity. An approach that unifies these different solutions is  Sakharov's induced gravity \cite{1968SPhD...12.1040S,2000GReGr..32..365S}. In this proposal, the quantum effects of the particles and fields in the universe give rise to the cosmological constant and gravitation itself. The quantum fluctuations give rise to the phase transition in gravity when the Ricci scalar $R$ becomes of order the mass square of the scalar field and freezes $R$  there. This model is interesting in terms of its physical origin and its nearly first principles derivation \cite{1999PhRvD..60l3502P,2006PhRvD..73b3513C}. The critical condition is given by

\begin{figure}
\includegraphics[width=.8\textwidth, height=7.5cm]{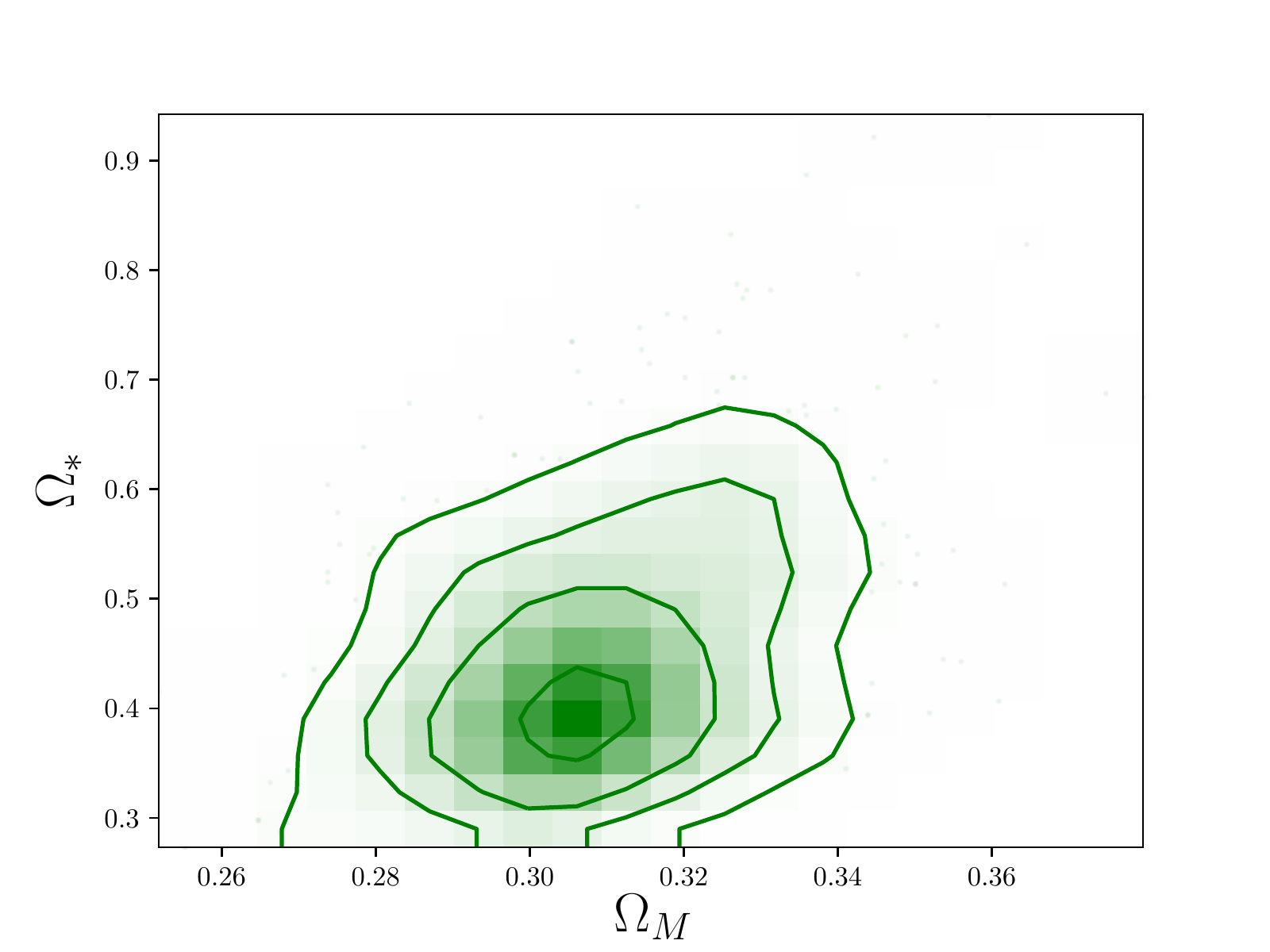}
\caption{Constraints on \om\, - \ost\, for the vacuum metamorphosis model. The best fit parameters are \om\,= 0.307 $\pm$ 0.018 and \ost = 0.44 $\pm$ 0.093 implying a possible non-zero $z_t$ at a modest 1.5 $\sigma$ level.}
\label{fig:vm}
\end{figure}

\begin{equation}
R = 6(\dot{H} + 2H^2) = m^2
\label{eq:crit_cond}
\end{equation}

\begin{equation}
\frac{H^2}{H_0^2} = (1 - \frac{m^2}{12})(1+z)^4  +\frac{m^2}{12}  , z < z_t 
\label{eq:hz_vm_pretrans}
\end{equation}

\begin{equation}
\frac{H^2}{H_0^2} = \Omega_M (1+z)^3 + \frac{m^2}{3} \frac{1 - \Omega_{*}}{4 - 3 \Omega_*} , z > z_t 
\label{eq:hz_vm_posttrans}
\end{equation}
where \ost\, is the matter density at the redshift of transition [$\Omega_m (z_t)$] and 1 - \ost\, behaves like the cosmological constant for redshifts greater than the transition redshift. 
$z_t$ and $m^2$ are given in terms of the density parameters by

\begin{equation}
z_t = \left(\frac{m^2\Omega_*}{3\Omega_M(4 - 3\Omega_*)}\right)^{1/3} - 1
\label{eq:transition_vm}
\end{equation}

\begin{equation}
m^2 = 3\Omega_M \left[\frac{4 - \Omega_*}{\Omega_*}\right]^{1/4} \left[\frac{4}{m^2} - 1/3\right]^{-3/4}
\label{eq:msq_vm}
\end{equation}
Note that $z_t = 0$ corresponds to standard \lam CDM.  
Figure \ref{fig:vm} presents the constraints for \om-\ost\, from the datasets in our study.
We obtain best fit values of \om\,= 0.307 $\pm$ 0.018 and \ost = 0.44 $\pm$ 0.093 . The marginalised constraints for each parameter are broadly consistent with \om=\ost. 
For the best fit values, we obtain a transition redshift z$_t =$ 0.20 $\pm$ 0.14 which only indicates a vacuum phase transition at a modest 1.5 $\sigma$ level. 

\subsection{Thawing Scalar Field Models}
Since a significant number of models tested in this study are motivated by the thawing scalar field, we present a background to scalar field cosmologies. 

The dynamical equations of motion for the scalar field is given by the Klein-Gordon equation

\begin{equation}
3 H M_{\mathrm{pl}}^2 = \rho_\phi + \rho_m
\label{eq:Eom1}
\end{equation}

\begin{equation}
\ddot{\phi} + 3 H \dot{\phi} + V_{,\phi} = 0
\label{eq:kg_equation}
\end{equation}
where the second term is a friction term commonly known as the ``Hubble drag''.

\begin{equation}
\dot{\rho_m} + 3 H \rho_m = 0
\label{eq:Eom2}
\end{equation}
where $H = \dot{a}/a$. $V(\phi)$ is the potential of the scalar field $\phi$ and $V_{,\phi} = \frac{dV}{d\phi}$.  The equation of state $w$ is given as the ratio of the pressure ($P$) to the density ($\rho$) where $P = \dot{\phi}^2/2 - V(\phi) $ and $\rho = {\phi}^2/2 + V(\phi)$ and the dimensionless density of the scalar field is \ophi = $\rho_{\phi}/(3 H^2 M_{pl}^2)$. 

From equations \ref{eq:Eom1},~\ref{eq:kg_equation},~\ref{eq:Eom2}, we can obtain the following expressions for $w$\, and \ophi\, (\cite{2008PhRvD..78l3525D,2008PhRvD..77h3515S} see also \cite{ 2015PhRvD..91f3006L})

\begin{equation}
w^{'} = (1-w) (-3(1+w) + \lambda \sqrt{3(1+w) \Omega_\phi}),
\label{eq:eos}
\end{equation}

\begin{equation}
\Omega_\phi^{'} = -3 w \Omega_\phi  (1 - \Omega_\phi),
\label{eq:ophi}
\end{equation}

\begin{equation}
\lambda^{'} = - \sqrt{3 (1+w) \Omega_\phi} (\Gamma - 1) \lambda^2,
\label{eq:lambda}
\end{equation}
where $\lambda = M_{pl} V_{,\phi}/V$ and $\Gamma = V V^{,\phi\phi}/V^{2}_{,\phi}$ and prime refers to a derivative with respect to $\mathrm{ln}\, a$.
In this study, the different scalar fields are motivated by thawing models, for which  the field is frozen in the matter-dominated era  with $w = -1$ due to Hubble friction and is thawing to less negative values in the recent past. For all models in this study we present an analytic equation of state which can be integrated to obtain the  \hz\,  and hence the distance measurements for obtaining the parameter constraints.  For all scalar field models we assume flatness, i.e. \om\, + \ophi\, = 1. We now proceed to present parameter constraints on different thawing models.

\subsubsection{Doomsday Model}
\label{ssec:doom}

A simple generalisation of the cosmological constant model is the linear potential model, pioneered by \cite{1987PhT....40i..61L} and discussed more recently in the literature by \cite{2008cosm.book.....W}. Interestingly,  while the linear potential  gives a current accelerating epoch, in the future, the potential becomes negative and not only deceleration of the expansion but collapse of the universe ensues. For this reason, the linear potential model is often referred to as the ``doomsday model''.

\begin{figure}
\includegraphics[width=.8\textwidth,height=7.5cm]{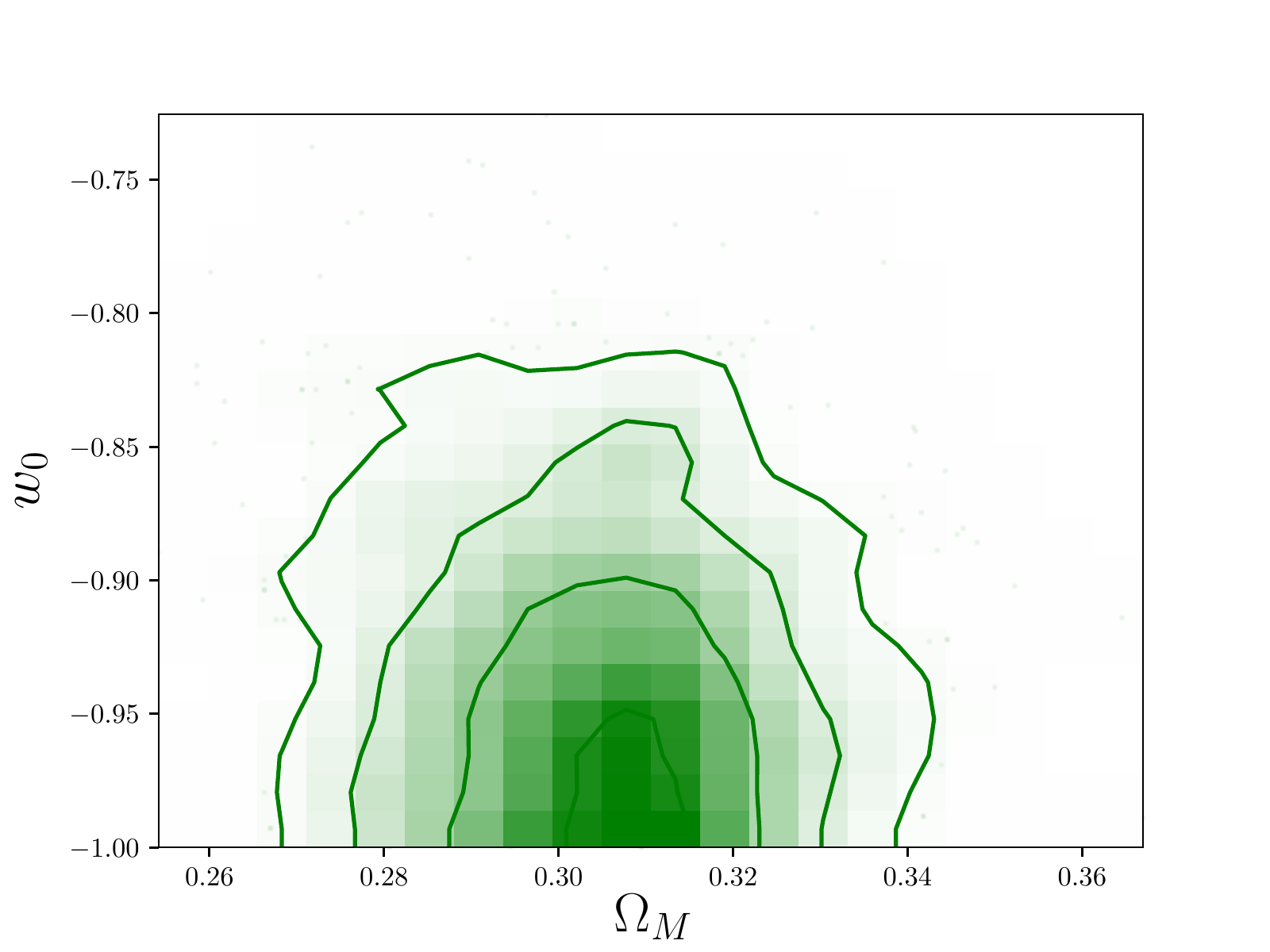}
\caption{Constraints on the present day matter density, \om\, and dark energy EoS \wo\, for the doomsday potential. The equation of state is constrained at 95 $\%$ C.L. to -0.82}
\label{fig:doom}
\end{figure}

The linear potential has two parameters, the amplitude $V_0$ and the slope $V^{'}=dV/d\phi$. The amplitude, $V_0$ corresponds to the present day dark energy density, which is related to \om\, in a flat universe (for all models in the remainder of the paper, we assume flatness).  The slope $V^{'}$ is indicative of the present day dark energy EoS, \wo. \cite{2003JCAP...10..015K} discuss the cosmological properties of the linear potential and \cite{1987PhT....40i..61L} present a view of it as a perturbation about a zero cosmological constant. This model has been considered a textbook case by \cite{2008cosm.book.....W}. Such a dark energy model is an example of a thawing scalar field \cite{2005PhRvL..95n1301C} wherein the dark energy EoS is -1 in the early universe (z $\gg$ 1) and slowly rolls to less negative values. If it hasn't evolved too far from -1, the EoS can be approximated as $w_a \sim -1.5 (1+w_0)$ \cite[see][]{2009ApJ...695..391R} where $w(a) = w_0 + w_a(1-a)$ \cite{2001IJMPD..10..213C,2003PhRvL..90i1301L}.

In \cite{2008cosm.book.....W}, the potential associated with the thawing scalar field is written as $V(\phi) = V_0 + V^{'}_0 (\phi - \phi_0)$ with $V_0$ being the potential energy in the initial frozen state at high redshift (with high Hubble drag) and $V^{'}_{0}$ is the constant potential slope. 
The constraints on the cosmological parameters are presented in Figure~\ref{fig:doom}. \wo \textless -0.82 at 95 $\%$ C.L for the SN data combined with the CMB/BAO ratio. 

\subsubsection{One parameter Slow-Roll Dark Energy}
Recently, \cite{2011MNRAS.416..907G,2014MNRAS.438.1948S} suggest that the simplest dark energy model has the same explanation as inflation, likely a scalar field slowly rolling down its potential, analogous to the one predicted in chaotic inflation \cite{1983PhLB..129..177L}. In such a model, dark energy will have a generic EoS and the universe will have a generic dependence of the Hubble parameter on redshift, independent of the potential's starting value and shape. The EoS and 
Hubble parameter offer the desired model independent, but physically motivated parametrisation, because they will hold up for most of the standard scalar field quintessence and phantom DE models. A particularly interesting feature of this model is that the dynamics of dark energy are described by a single parameter instead of a two parameter prescription \cite[for e.g.][]{2008PhRvD..78l3525D,2009PhRvD..79h3517C,2013PhRvD..87h3505C}. 
The Hubble parameter for this model is given by
\begin{equation}
\frac{H^2}{H_0^2} = \Omega_M (1+z)^{3} + (1 - \Omega_{M}) \left[\frac{(1+z)^3}{\Omega_M(1+z)^3 + 1 - \Omega_M}\right]^{\delta w_0/(1 - \Omega_M)}
\label{eq:hz_slowroll}
\end{equation}
which assumes flatness and $\delta w_0 = 1 + w $ \cite{2014MNRAS.438.1948S}. 	
The best fit constraints on \dw\, are -0.008 $\pm$  0.107 which is consistent with the cosmological constant, but allows for small departures from $w_0 = -1$. Our constraints on \dw\, are slightly less precise than the recent \emph{Planck} \cite[\dw\, = -0.008 $\pm$ 0.068;][]{2016A&A...594A..14P} and BAO estimates \cite[\dw\,=0.05 $\pm$ 0.07][]{2015PhRvD..92l3516A} as we use the CMB/BAO ratio which has larger uncertainties than either the CMB compressed likelihood ($\sim$ 0.5 $\%$ on the CMB shift, $R$, for e.g.) or the BAO angular scale. Since the critical parameter in this model is the present day EoS of dark energy and not the EoS at a pivot redshift ($w_p$), low-redshift probes like SNe would be extremely useful in determining departures from a cosmological constant, which has also been demonstrated in forecasts by \cite{2014MNRAS.438.1948S}. 

\begin{figure}
\includegraphics[width=.8\textwidth, height=7.5cm]{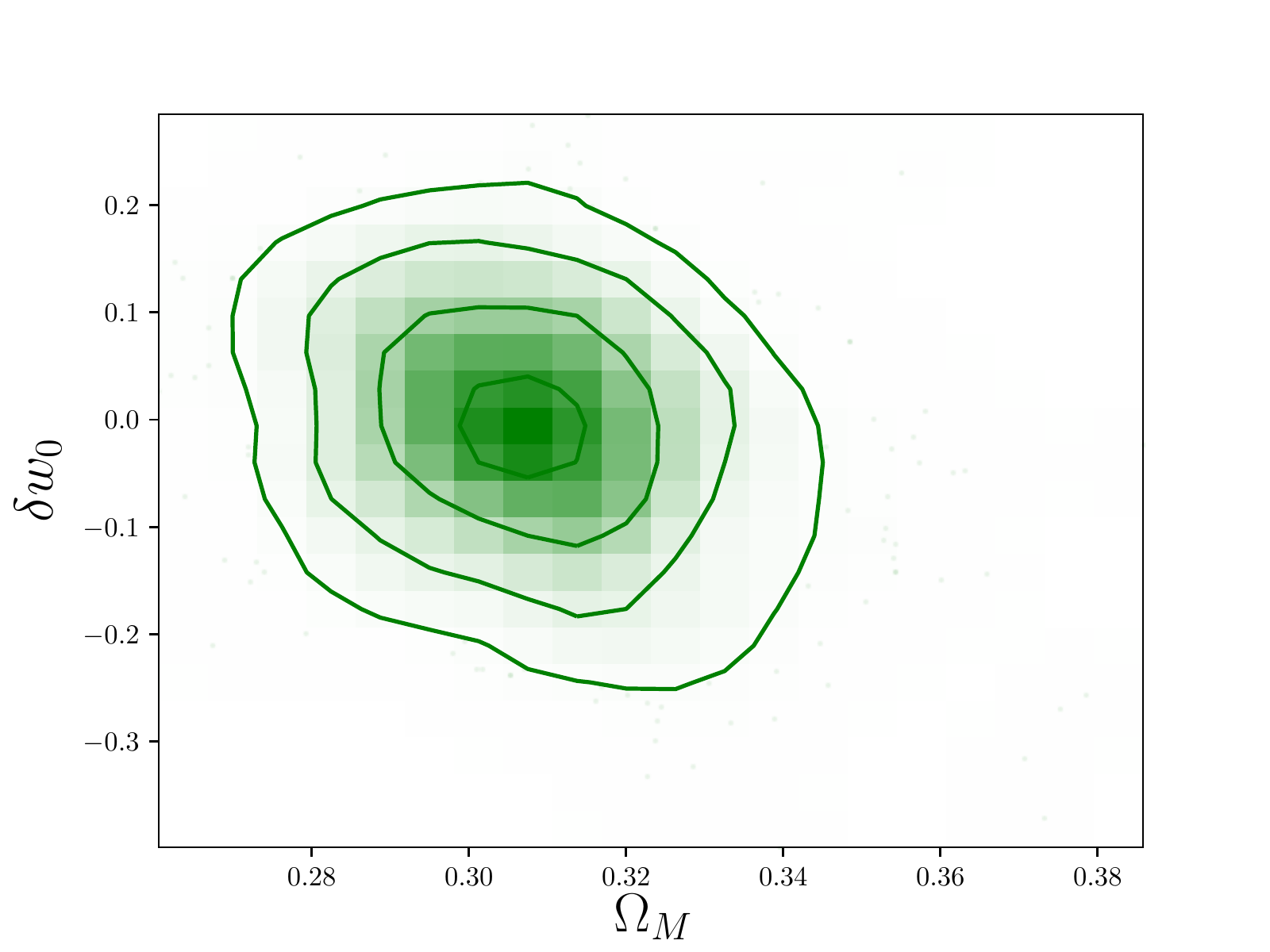}
\caption{Marginalised constraints on the matter density \om\, \dw. \dw\, = -0.008 $\pm$ 0.107 which is consistent with a cosmological constant but has some room for deviation. The constraints are slightly weaker than the recent \pl\, values \cite{2016A&A...594A..14P} }
\label{fig:srde_oneparam}
\end{figure}

We note that the best fit value for \dw\, using the \pl\, compressed likelihood and BAO $d_z$ is 0.012 $\pm$ 0.070 which is consistent with the estimates in \cite{2015PhRvD..92l3516A,2016A&A...594A..14P} and with the constraints from the CMB/BAO ratio, but with a smaller error.

\subsubsection{Pseudo-Nambu-Goldstone-Boson}
One of the key puzzles is how to prevent quantum corrections from adding a Planck energy scale cosmological constant or affecting the shape of the potential, known as the issue of technical naturalness. Pseudo-Nambu-Goldstone-Boson (PNGB) models are technically natural due to a shift symmetry and so can be considered well motivated. 
\cite{1995PhRvL..75.2077F} present an early cosmological analysis of dark energy as PNGB with more updated works by \cite{2007PhRvD..75f3514D, 2008PhRvD..77j3503A, 2017JCAP...01..023S}

The potential for the model is

\begin{equation}
V(\phi) = V_*  [cos(\phi/f) + 1], 
\label{eq:pngb_potential}
\end{equation}
with $V_*$ setting the amplitude, $f$ the symmetry scale and the initial value of the field when it thaws from the high redshift, high Hubble drag, frozen state $\phi_i$. These parameters can be thought of as roughly analogous to the dark energy density, dark energy EoS and the time variation of the EoS. 
PNGB models are an example of thawing dark energy where the field has recent departed from its high redshift cosmological constant behaviour, evolving towards a less negative EoS.  In the future, the field oscillates around its minimum with zero potential and ceases to accelerate the expansion, behaving as non-relativistic matter. Since the deviation from $w_0\,= \, -1$ is recent, the precision in measuring \wo\, is more important that measuring the averaged or pivot EoS value. This implies that probes like SN~Ia provide the tighest constraints on \wo.

An approximate equation of state with the associated scalar field potential was calculated by \cite{2008PhRvD..78l3525D} and \cite{2009PhRvD..79h3517C} and analysed in context of observational data in \cite{2013PhRvD..87h3505C}. Our main new contribution to the parameter estimation is newer data along with a model independent approach for using the CMB and BAO data. For a derivation of the parameters for $V(\phi)$ see \cite[][]{2017JCAP...01..023S} where 

\begin{figure*}
\begin{equation}
w(a) = -1 + (1 + w_0)^{3a(K-1)} \cdot \left[ \frac{(K - F(a))(F(a) +1) + (K+F(a))(F(a)-1)]}{(K - \phiroot)(\phiroot+1)^K+(K+\phiroot)(\phiroot - 1)^K} \right] ^2
\label{eq:pngb_eos_approx}
\end{equation}
\end{figure*}

\begin{equation}
K = \sqrt{1 - \frac{4M_{\mathrm{pl}}^2 V_{,\phi\phi}(\phi_i)}{3V(\phi_i)}}
\label{eq:k_pngb}
\end{equation}

\begin{equation}
F(a) = \sqrt{1 + (\Omega_\phi^{-1} - 1)a^{-3} }
\label{eq:fa_pngb}
\end{equation}
where $F(a)$ is the inverse square root of the fractional energy density corresponding to the cosmological constant. 

\begin{figure}
\includegraphics[width=.5\textwidth, trim = 0 10 0 30]{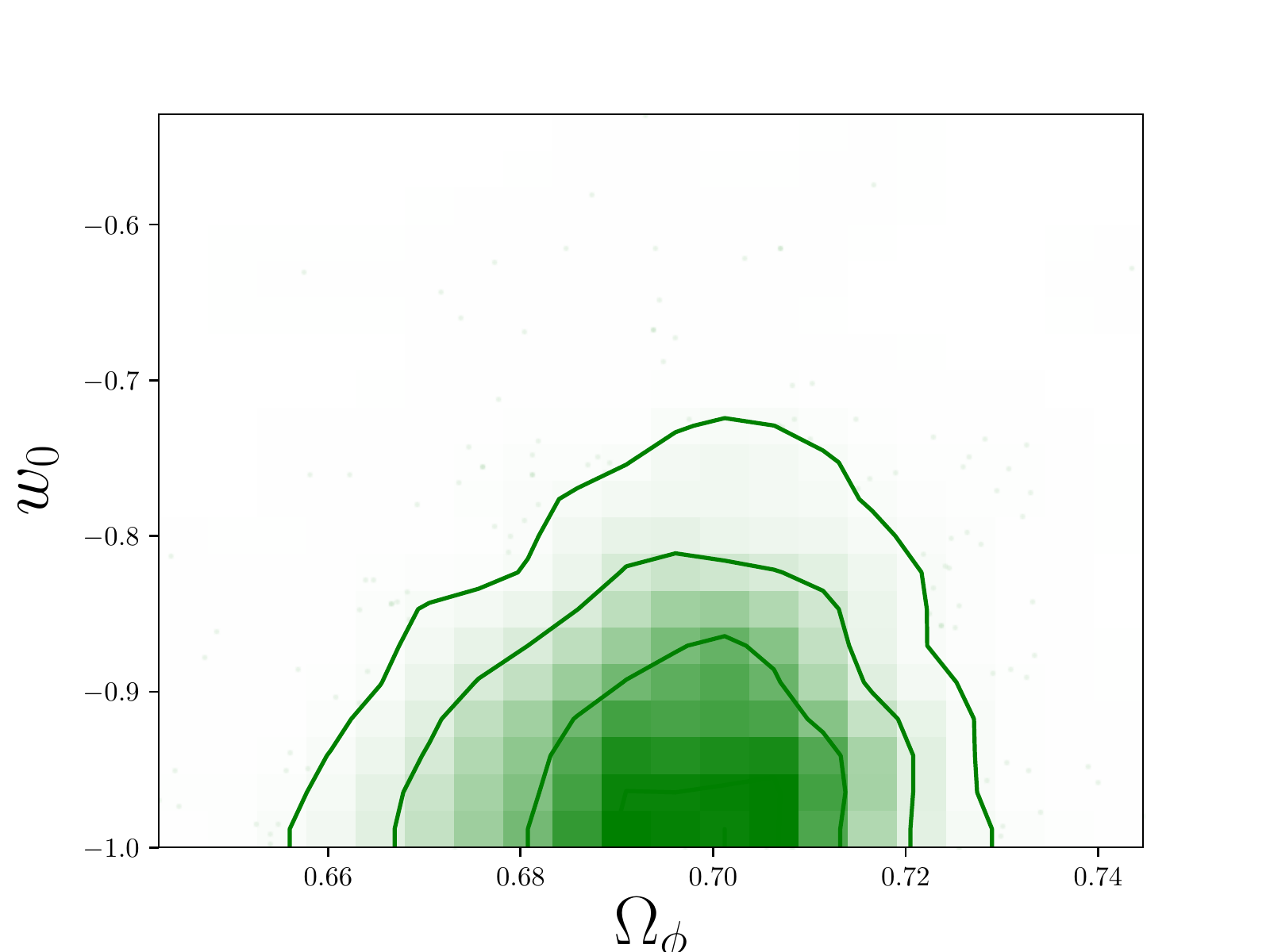}
\includegraphics[width=.5\textwidth, trim = 0 10 0 30]{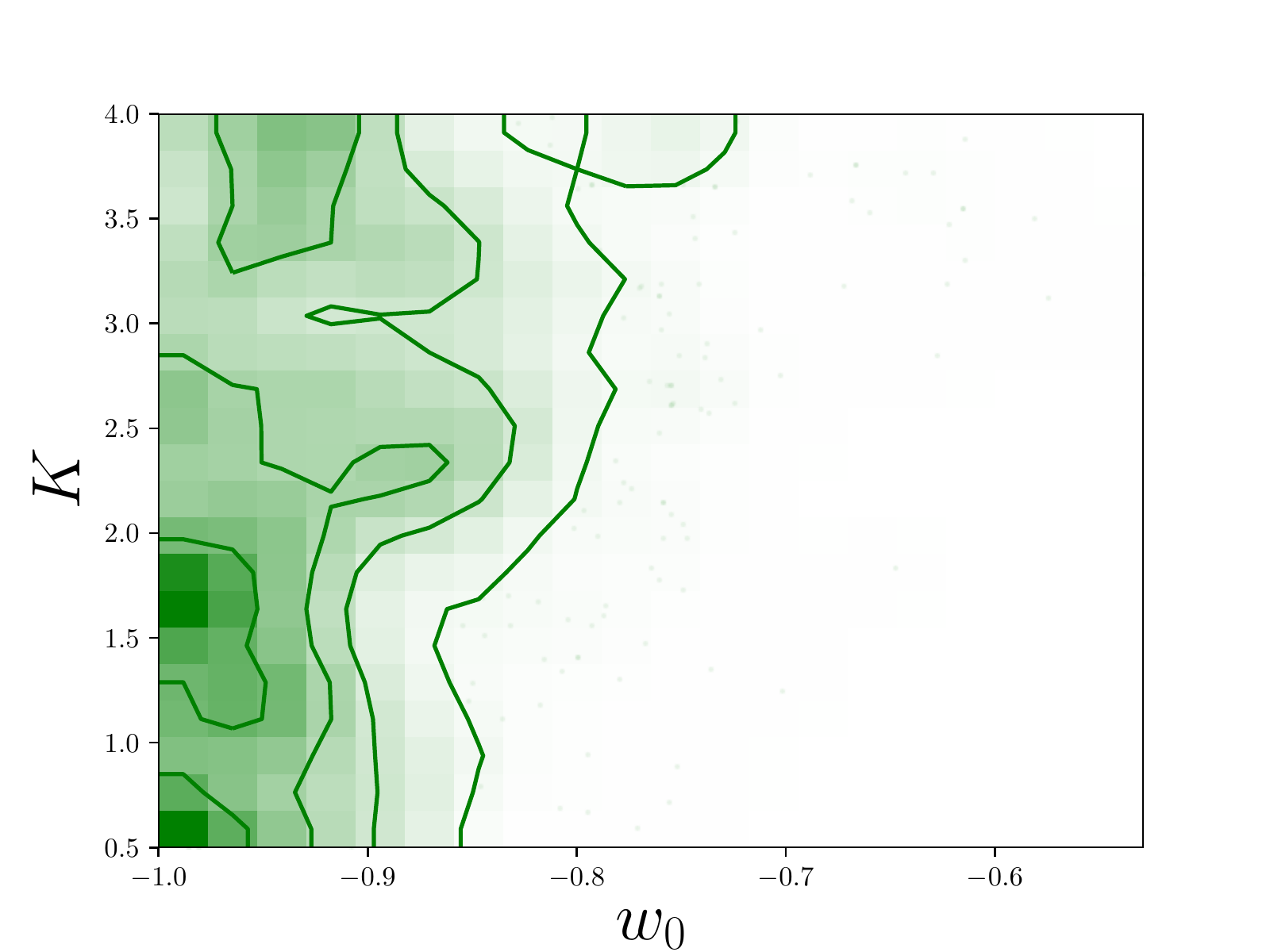}
\caption{(Left): Constraints on \ophi\, and \wo\, from SN~Ia combined with the CMB/BAO ratio (Right):Constraints on the present day equation of state, \wo\, and the curvature of the potential, as given by equation~\ref{eq:k_pngb}}
\label{fig:pngb}
\end{figure}
We derive luminosity distances for this model by using equation~\ref{eq:pngb_eos_approx} in general expression for $\Omega_{\mathrm{DE}}(z)$ in equation~\ref{eq:ode_z}
where \ophi\, is the present day energy density of the field, \wo\, is the present day equation of state and $K$ measures the curvature of the potential at maximum \cite{2013CQGra..30u4003T}. For $K$ \textgreater 10, the motion of the field at the start of the evolution is required to be very small to avoid a quick roll down. If the field touches the minimum potential and starts oscillating at a scale factor value near today's, numerical simulations establish that equation ~\ref{eq:pngb_eos_approx} is not valid anymore. In addition to this inaccuracy, for an oscillating potential, the equation of state would become positive, therefore violating the $w$ \textless -1/3 condition for dark energy description.
For $K < 0.5$, the field mass becomes very large, implying that the Taylor expansion around $\phi = \phi_i$ is invalid due to the rapid variation of the field. Hence, this value sets the lower bound on the prior for $K$.

Applying a prior of \wo\, $\geq -1$ for the present day equation of state (characteristic of a thawing quintessence model) we derive constraints on the three parameters in the model, namely, \ophi, \wo\, and $K$. The constraints for \wo\, and $K$ are presented in Figure~\ref{fig:pngb}. \wo\, \textless -0.75 and 0.66 \textless \, \ophi\, \textless 0.73 at the 95$\%$ C.L.
\cite{2009PhRvD..79j3005D} argue that phantom fields can be accommodated within equation~\ref{eq:pngb_eos_approx}. The resulting constraints on \wo\, \ophi\, and $K$ are shown in Figure~\ref{fig:pngb}.

Previously \cite{2013PhRvD..87h3505C} presented constraints on the model parameters for the PNGB approximate EoS. The authors, however, use a slightly different formalism for the input dataset. They use the BAO $A(z)$ parameter \cite{2005ApJ...633..560E} along with a compressed CMB likelihood 



\cite{2017JCAP...01..023S} present constraints on the parameters of the PNGB potential using the latest CMB, SN~Ia and BAO data. They compare their results to the approximate equation of state, however, they use priors on $K$ calculated from the numerical solution of the background evolution equation and hence, obtain a stricter posterior distribution of $K = 1.1 \pm 0.4$.

\subsubsection{Algebraic Thawing}
PNGB models require a pseudoscalar thawing field. However, we can also consider scalar fields with a thawing behaviour. Any such fields that are neither fine-tuned nor have overly steep potentials must initially depart from the cosmological constant behaviour along a specific track in the EoS phase space, characterised by the form of a slow roll behaviour in the matter-dominated era \cite[for e.g., see][]{2008PhRvD..77h3515S,2008JCAP...11..015C}

In this study we adopt the model of \cite{2008GReGr..40..329L} designed to incorporate the physical behaviour given by the following EoS

\begin{equation}
1 + w = (1+ w_0) (1+z)^{-p} \left( \frac{1+b}{1+b(1+z)3} \right) ^{1 - p/3} 
\label{eq:eos_alg_thaw}
\end{equation}
implying a normalised Hubble parameter given by
\begin{equation}
\frac{H^2}{H_0^2} = \Omega_M (1+z)^3 + (1 - \Omega_M) \times \\
\mathrm{exp} \left[ \frac{3(1+w_0)}{\alpha p} \{1 - (1 - \alpha + \frac{\alpha}{(1+z)^3})^{p/3} \}
\right]
\label{eq:hz_alg_thaw}
\end{equation}
where $\alpha = 1/(1+b)$ and $b = 0.3$ is a fixed constant and not a parameter. Two parameters (apart from the present day matter density \om) are \wo\, and $p$. This form follows the scalar field dynamics not only to leading order but also to next-to-leading order \cite[see][]{2008JCAP...11..015C}

\begin{figure}
\centering
\includegraphics[width=.8\textwidth, height=7.5cm]{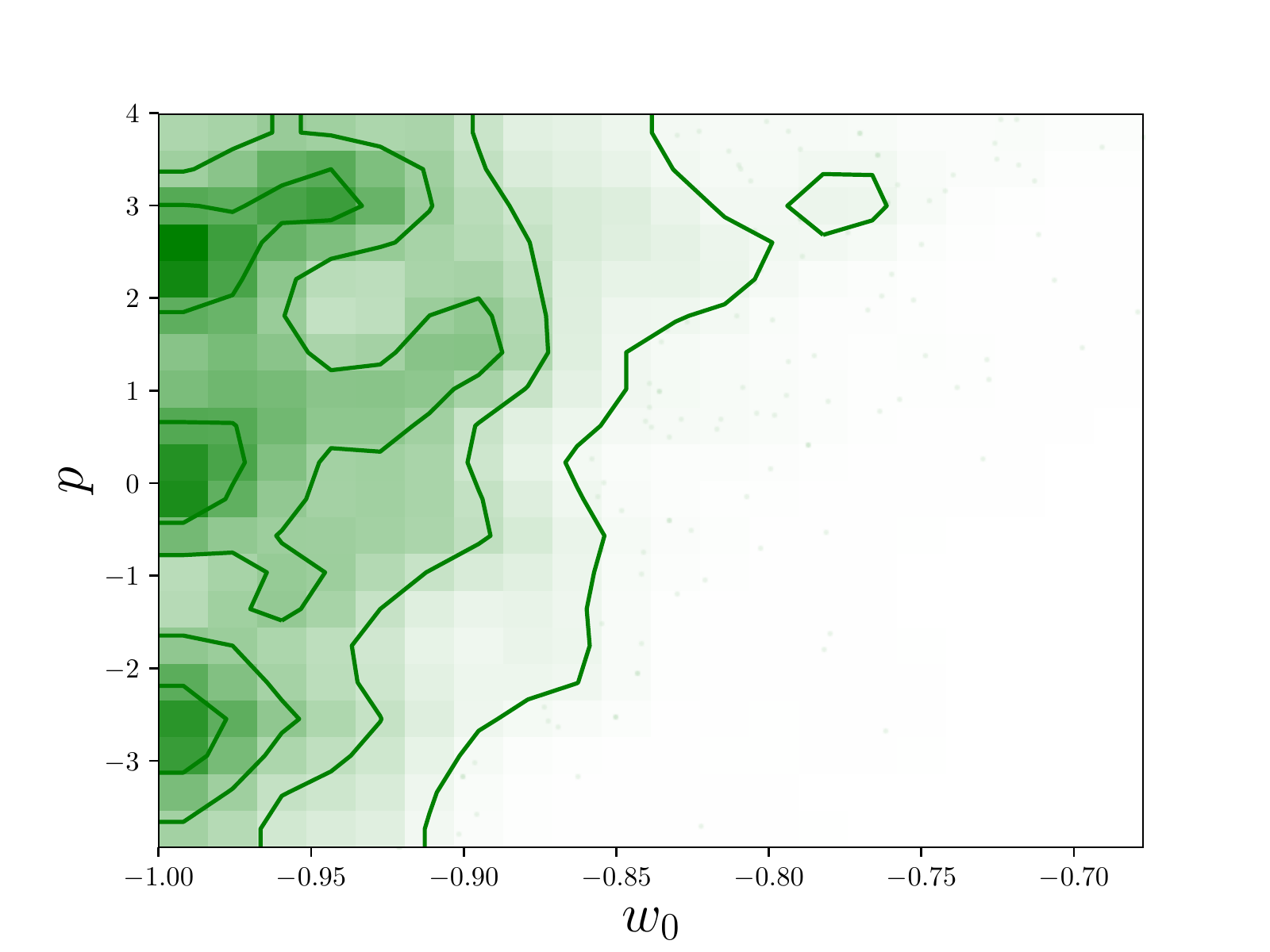}
\caption{Constraints on the present day equation of state ($w_0$) and power law exponent, $p$ for algebraic thawing model given in equation~\ref{eq:hz_alg_thaw}. The 95$\%$ C.L. constraints on \wo\, from SNe and the CMB/BAO ratio are \textless -0.78.}
\label{fig:alg_thaw}
\end{figure}

The physical behaviour of a minimally coupled scalar field evolving from a matter-dominated era would tend to have $p \in$ [0,3]. Hence, to test whether the data point to such a thawing model we consider $p$ values outside this range. Results are shown in Figure~\ref{fig:alg_thaw}, the current data imply \wo \textless -0.78. This estimate is consistent with the other thawing quintessence models tested here, namely, doomsday and PNGB cases. 

We test a more updated algebraic thawing model of \cite{2015PhRvD..91f3006L} in which $p = 1$, hence, it is only a two parameter description of the \hz. In this case $b = 0.5$, which describes the behaviour of the field in the matter dominated era. For this model, we obtain \wo\, \textless -0.

\subsection{Growing Neutrino Mass}
\label{ssec:grownu}

We have studied the thawing class of quintessence models. Freezing or tracker models \cite[in which the dark energy EoS is less negative in the early universe][]{2017arXiv170101445L} are interesting from the point of view of a fundamental physics since they help solve the fine-tuning problem due to the attractor behaviour in their dynamics, drawing from a large basin of attraction in the initial conditions \cite{1999PhRvL..82..896Z}. However, despite the interesting physical motivation, in freezing quintessence, it is difficult to have a sufficiently negative EoS at the present day. 

This problem is solved in growing neutrino models, wherein the mass of the neutrino ($m_\nu$) increases with time and stops the dynamical evolution of the dark energy scalar field solving the problem of the present day EoS. This forces the scalar field to have cosmological constant behaviour at the epoch when the neutrinos become non-relativistic, thus explaining the cosmological coincidence problem \cite{2004JCAP...10..005F,2007PhLB..655..201W}. 
Recent constraints on the mass varying neutrino cosmology with CMB and weak gravitational lensing (WL) probes are presented in \cite{2013A&A...560A..53L} as well as constraints from CMB alone in \cite{2014PhRvD..90d5002G}. The combined dark sector (scalar field plus neutrinos) energy density is given by 

\begin{equation}
\Omega_{\mathrm{ds}}(a) = \frac{\Omega_{\mathrm{ds}} (a^3) + 2\Omega_\nu (a^{3/2} - a^3)}{1 - \Omega_{\mathrm{ds}} (1 - a^3) + 2\Omega_\nu (a^{3/2} - a^3)}; \,\,  a \textgreater a_t \\
\label{eq:ods_grownu_posttrans}
\end{equation} 
\begin{equation}
\Omega_{\mathrm{ds}}(a) = \Omega_e; \,\,\, a \textless a_t 	
\label{eq:ods_grownu_pretrans}
\end{equation}
where \ods = 1 -\om\, is the present day dark energy density. The scale factor at which the neutrinos become non-relativistic, is given by preserving continuity between the early and late time terms (i.e. setting equations~\ref{eq:ods_grownu_pretrans} and ~\ref{eq:ods_grownu_posttrans} at $a = a_t$). The two free parameters are the early dark energy density \ome and the neutrino density \onu. 
The normalised Hubble parameter is given by 
\begin{equation}
\frac{H^2}{H_0^2} = \frac{\Omega_M a^{-3}}{1 - \Omega_{\mathrm{ds}}(a)},
\label{eq:hz_grownu}
\end{equation}
and the EoS 
\begin{equation}
w(a) = -1 + \frac{\Omega_{\nu} a^{-3/2}}{\Omega_{\mathrm{ds}} + 2 \Omega_\nu(a^{3/2}-1)}, a \textgreater a_t
\label{eq:grownu_eos}
\end{equation}
with $w = 0$ before the transition. Therefore, one can translate the \onu\, or $m_\nu$(z = 0) to an EoS (substituting $w$=\wo\ at $a=1$ in \ref{eq:grownu_eos}, $w_0 = -1 + \Omega_\nu/\Omega_{ds} = -1 + \Omega_\nu/(1 - \Omega_M)$).  

\begin{figure*}
\includegraphics[width=.8\textwidth, height=7.5cm]{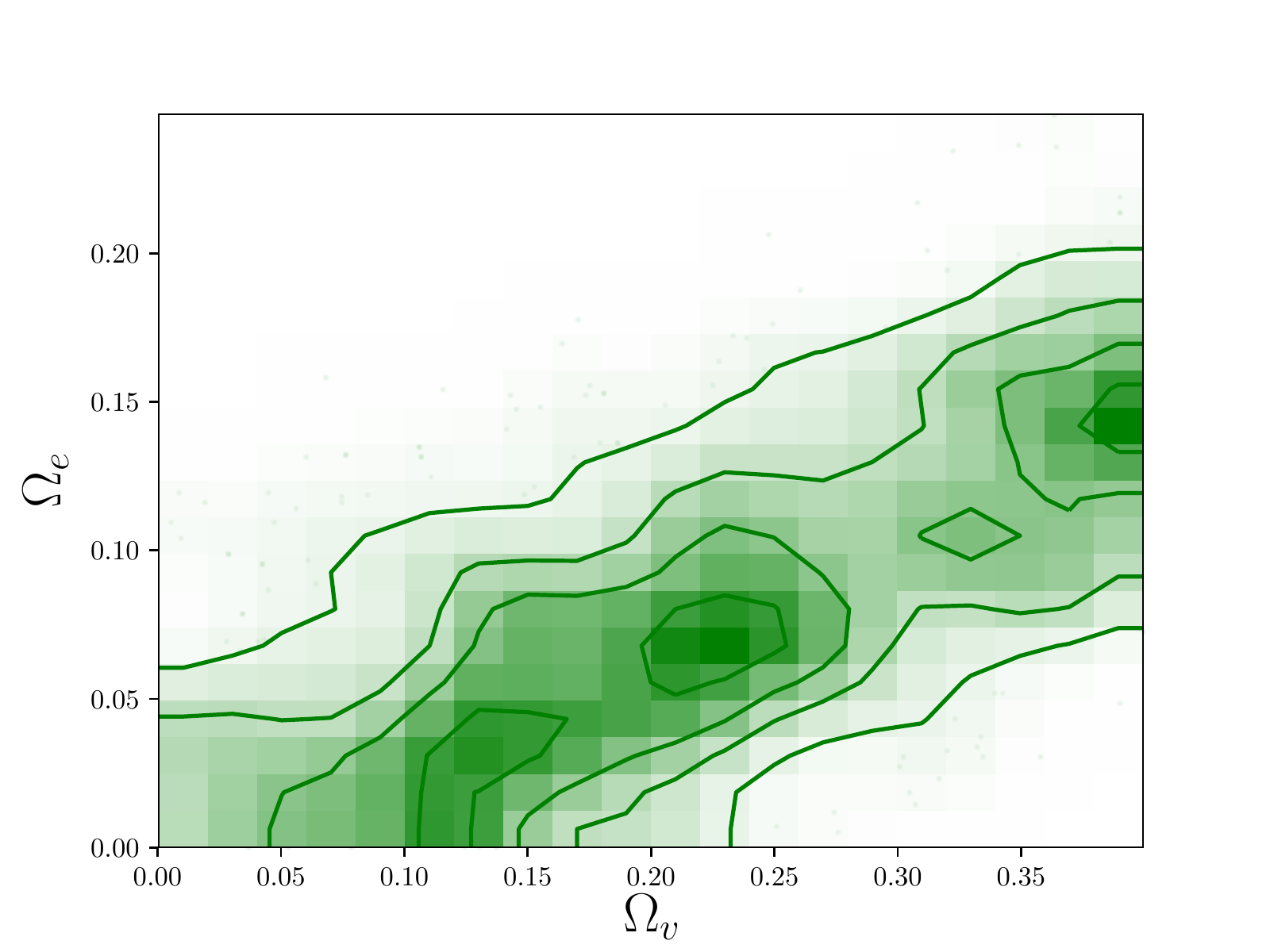}
\caption{Growing neutrino mass model, by coupling the dark energy scalar field to massive neutrinos can solve the cosmological coincidence problem. Constraints on \onu - \ome\, from SN~Ia combined with the CMB/BAO ratio.}  
\label{fig:grownu}
\end{figure*}

From equation~\ref{eq:hz_grownu} we can see that this model has an early component of dark energy. Hence, the compressed CMB likelihood defined in Section~\ref{ssec:cmb} would not be valid and we can only use the CMB/BAO ratio, $f$, for analysing the parameter constraints. The resulting contours are shown in~\ref{fig:grownu}

From Figure~\ref{fig:grownu}, we can see that there is a strong degeneracy between the \ome\, and \onu\, when using the CMB/BAO ratio. The constraint $m_{\nu}$ \textless 2 eV  is not very strict compared to the latest studies \cite[e.g.][]{2017arXiv170108172V}, which is expected since we don't use the complete CMB or any structure formation information. In a future work we aim to constrain the early dark energy and neutrino densities using the complete CMB and large scale structure data combined with SNe and BAO.

\begin{table*}
\caption{Priors on the free parameters for the models tested in this study. The four parameters listed here, namely, $\alpha$, $\beta$, $\Delta_M$, $M_B$ are nuisance parameters that need to fit to the supernova data. We use a flat prior on $H_0$}
\begin{tabular}{|l|c|c|}
\hline\hline
Model Parameter & Prior & Model\\
\hline
\om & U[0, 0.5] & All except PNGB\\
\ophi & U[0, 1] & PNGB\\
\ok & U[-0.5, 0.5] & \lam CDM\\
\ost & U[0, 1] & Vacuum Metamorphosis \\
\wo & U[-1, 1] & Doomsday, Algebraic thawing, PNGB\\
\dw & U[-2, 1] & One-parameter Slow-roll DE \\
$p$ & U[-4, 4] & Algebraic thawing \\
$K$ & U[0.5, 4] & PNGB  \\
\ome & U[0, 0.25] & Growing $\nu$ mass\\
\onu & U[0, 0.4] & Growing $\nu$ mass\\
$B_2$ & U[-100, 1.4] & Bimetric:Quadratic \\	
$H_0$ & U[50, 100] & Hubble constant, all models\\
&&\\
$\alpha$ & U[0, 1] & SN~Ia correction for light curve width\\
$\beta$ & U[0, 4] &SN~Ia correction for colour\\
$\Delta_M$ & U[0, 0.2] & SN~Ia correction for host galaxy mass\\
$M_B$ & U[-35, 15] & SN~Ia absolute magnitude \\

\hline
\end{tabular}
\label{tab:prior}
\end{table*}

\subsection{Bimetric Gravity}
\label{ssec:bimetric}
Early attempts to modify the equations of general relativity included introducing a mass to the theory, effectively giving mass to the  particle mediating the gravitational force, the graviton \cite{1939RSPSA.173..211F, 1971PhRvD...3..867I}. It was long believed that massive gravity theories necessarily contained fatal ghost modes \cite{1972PhRvD...6.3368B}. Recently, it was suggested that the inclusion of a second metric and a carefully constructed interaction between the two metrics of the theory could remove the ghost problem \cite{2011PhRvL.106w1101D}, proven in \cite{2012JHEP...02..026H,2012PhRvL.108d1101H}. Later work also allowed for dynamics of the second metric, so called bimetric gravity \cite{2012JHEP...02..126H}.

Here, we test bimetric gravity models where matter couples directly to only one of the metrics. The background cosmology from bimetric gravity includes an accelerated expansion term without requiring an explicit contribution from vacuum energy \cite{2012JCAP...03..042V,2012JHEP...01..035V,2012JHEP...03..067C,2013JHEP...03..099A}. Here we test the two simplest models with only linear, and linear and quadratic interactions.

\subsubsection{Linear Interaction}
Assuming a flat universe,  one can argue that the simplest bimetric models corresponds to including only the linear interaction terms between the two metrics. The corresponding \hz\, is given by

\begin{equation}
\frac{H^2}{H_0^2} = \frac{\Omega_M (1+z)^3}{2} + \sqrt{\left(\frac{(\Omega_M(1+z)^3) ^2}{2}\right)^2 + 1 - \Omega_M}
\label{eq:hz_bimetric_b1}
\end{equation}

We obtain \om\, of 0.334 $\pm$ 0.019. The linear interaction model fits the CMB/BAO ratio and SNe~Ia datasets well, when treated independently, however, the combined fit is poor, as reflected in the high $\chi^2$ in Table~\ref{tab:model_selec}. This result illustrates the importance of complementary geometric probes to exclude cosmological models. 

\subsubsection{Linear and Quadratic Interaction}
We extend the model to also include the quadratic interaction term, i.e. 
\begin{equation}
\frac{H^2}{H_0^2} = B_2 + \frac{B_1}{3r}
\label{eq:hz_b2}
\end{equation}
where 

\begin{equation}
B_1^2 = \frac{9 (1 - \Omega_M) (1 - B_2)^2}{3 - 2 B_2},
\label{eq:b1}
\end{equation}
and $r$ is the ratio of the scale factors of the $g$ and $f$ metrics. The evolution of $r$ is given by
\begin{equation}
B_2 r^3 + B_1 r^2 + (\Omega_M (1+z)^3 - B_2) r - B_1/3 = 0 ,
\label{eq:bimetric_r}
\end{equation}
The upper limit on the prior for $B_2$ is set by \om\, \textless 1. For arbitrarily negative values of $B_2$ this model approaches the standard \lam CDM case

The effective dark energy density $\Omega_{\mathrm{DE}}$ is given by $r (B_1 + B_2 r)$.

\begin{figure}
\includegraphics[width=.8\textwidth, height=7.5cm]{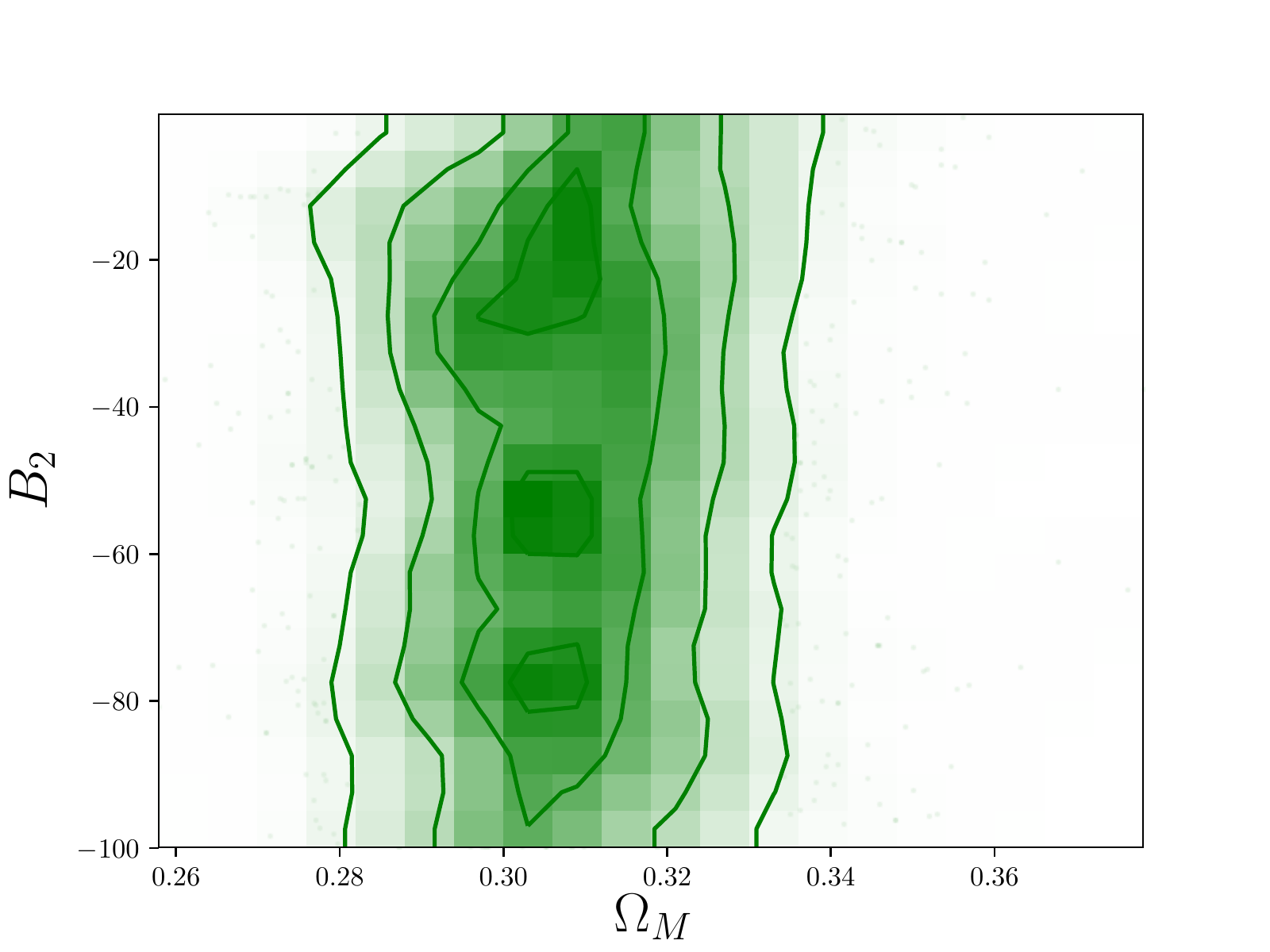}
\caption{Constraints on \om\, and $B_2$ in the bimetric gravity single coupling model with \om\, and $B_2$. \om\, is constrained tightly by the combination of SNe and CMB/BAO but allowed $B_2$ goes to infinitely negatively values, approaching the standard \lam CDM scenario}
\label{fig:b2_bimetric}
\end{figure}

\section{Model Selection}
\label{sec:mod_sel}

The estimates presented in Section~\ref{sec:param_est}  are statistical measures of model parameters assuming that the underlying model is the correct one. 
Parameter estimates along with model selection criteria can help us decide between possible models that explain the data. The $\chi^2$ statistics can be used to test the validity of a model but comparing relative likelihoods based on the $\chi^2$ doesn't account for the structural differences between them. In other words, $\chi^2$ values are useful for finding the best parameters in a model but are insufficient for deciding whether the model itself is the best one. 
Thus, in this study we turn towards a model selection criterion of the Bayesian evidence \cite[$Z$;][]{2004AIPC..735..395S} which is defined as the integral of the likelihood over the prior. 
\begin{equation}
Z = \int L \pi d \theta
\label{eq:evidence}
\end{equation}
where $L$ is the likelihood, $\pi$ is the prior and $\theta$ is the set of the parameters.  Moreover, with the exception of the bimetric gravity model with only the linear interaction term, all models fit the data well, hence, the $\chi^2$ doesn't distinguish the different models. 

Compared to simpler prescriptions for information criteria (e.g. Akaike Information Criterion, Bayesian Information Criterion \cite{1974ITAC...19..716A,1978AnSta...6..461S}) which only account for the number of parameters in a model, the Bayesian evidence can account for how much the allowed volume in data space increases due to the addition of extra parameters (i.e. how much more flexible the model becomes), as well as correlations between the parameters. \cite{2004MNRAS.348..603S} pioneered the use of Bayesian evidence in cosmology and \cite{2006PhRvD..74l3506L} used it compare different parametrisations of $w$, finding that the standard cosmological constant remains the favoured model.

In this paper, we calculate the evidence using a multimodal nested sampling algorithm called \texttt{MultiNest} \cite{2008MNRAS.384..449F, 2009MNRAS.398.1601F,2013arXiv1306.2144F} using a 
\texttt{python} module \cite{2014A&A...564A.125B}.  The model comparison statistic is the logarithm of the Bayes Factor (ratio of the Bayesian evidence for the two models being compared). 
The strength of the evidence for preferring one model over the other is given in Table~\ref{tab:dlogZ}. A complete discussion of this scale is presented in \cite{2008ConPh..49...71T}. 




For our model comparison, we firstly calculate the evidence for baseline \lam CDM cosmology, both with an assumption of flatness, as well as with curvature as a free parameter. For all non-standard models, except vacuum metamorphosis, we assume flatness, hence, the comparison of the evidence for all these models should be with the flat \lam CDM case. We find that three of the thawing models tested here, namely, the doomsday model, PNGB and algebraic thawing have a \dlz\, relative to flat \lam CDM \textgreater 2.5 ($\sim$ 3) which is significant evidence that the data prefer a flat \lam CDM over the more exotic cosmologies (in ratio of evidences is 20:1 in favour of flat \lam CDM). This is the first instance of strong evidence against these different thawing quintessence scenarios. Note that the doomsday model only has a single additional parameter, the present day EoS (however, the other two models have a total of 3 parameters, 2 more than flat \lam CDM). The one-parameter slow-roll dark energy, however, fairs better with a 
\dlz\, of  - 1.8, which is only weak evidence against this model relative to \lam CDM. The growing neutrino mass model has nearly the same evidence value as flat \lam CDM despite two additional parameters in the model fit. Finally, for the vacuum metamorphosis model, the \dlz\, relative to flat \lam CDM is only -1.5 which is weak evidence to exclude this model. However, since there isn't an assumption of flatness for this model, we compare it to standard \lam CDM where the \dlz\, is 1.1, suggesting a preference of vacuum metamorphosis compared to \lam CDM. 
The bimetric gravity model with linear interaction fares poorly relative to the standard case (\dlz\, =  -4.0), however the addition of the quadratic interaction helps improve the fit such the the evidence ratio is below the `moderate to strong threshold'. Moreover, we note that keeping curvature density \ok\, as a free parameter in \lam CDM cosmology is moderately disfavoured relative to the flat \lam CDM case. 

For the exotic cosmological models tested here we use a prior region for the entire theoretically viable parameter space (see Table~\ref{tab:prior}).We also test the effect of further relaxing the prior region on the parameters in the different models. As a test case we relax the \wo\, prior (for the algebraic thawing case, although this can be applied to any of the other models) to U(-1, 4.), for which the \dlz\, value decreases by 0.3. Hence, we find that using a less restrictive prior would further increase the model selection criterion favouring the standard cosmology. Similarly, increase the allowed prior for \ok\, in the \lam CDM model leads to a lower evidence relative to the flat \lam CDM case. Interestingly, for the bimetric gravity model considering the linear and quadratic interaction terms, extending the prior to more negative values for the $B_2$ parameter leads to a higher evidence. This is because the model tends towards \lam CDM for $B_2 \rightarrow \infty$.

\begin{figure}
\includegraphics[width=.9\textwidth]{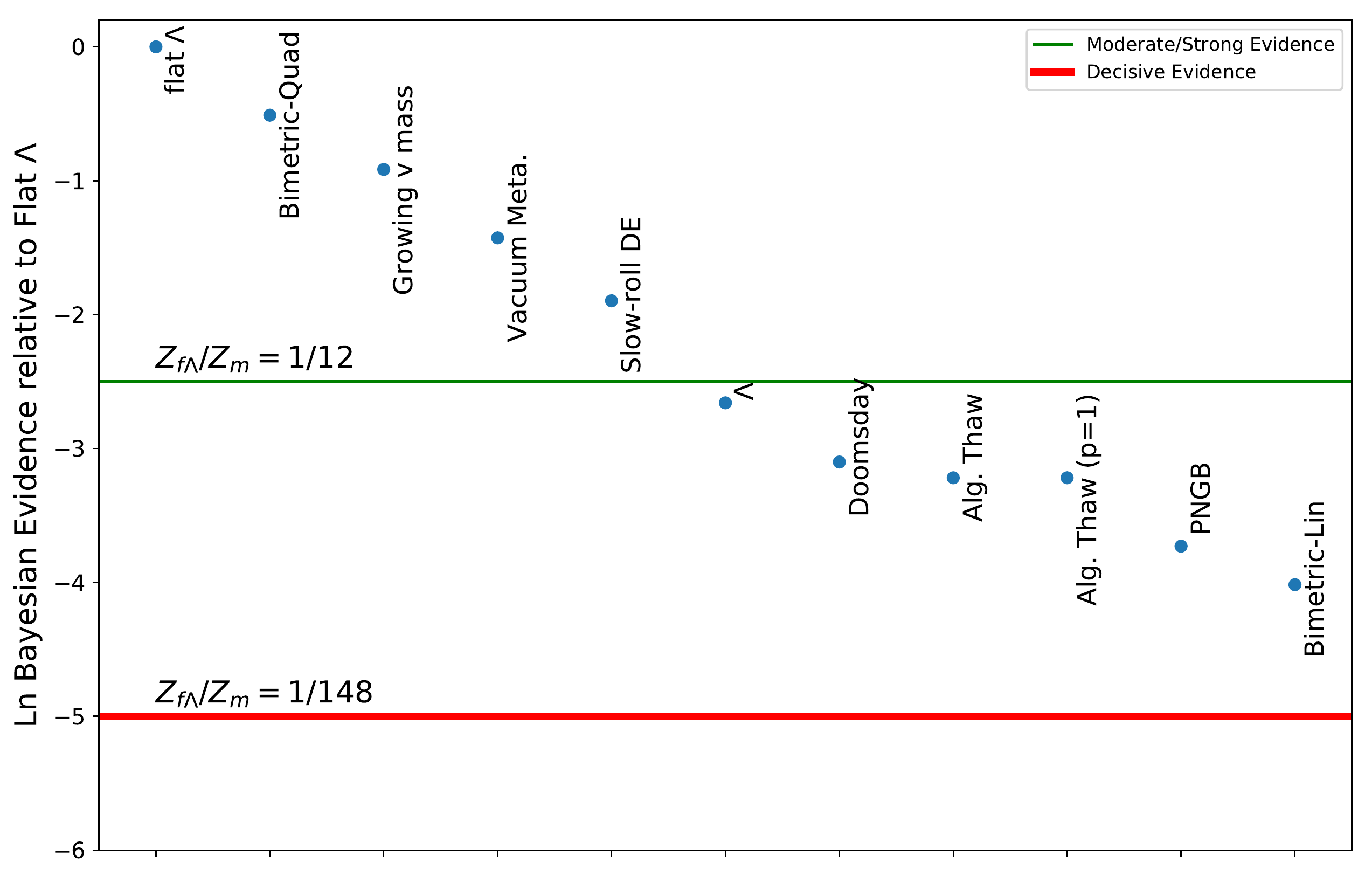}
\caption{The log of the evidence for flat \lam CDM to the evidence for the different non-standard models tested in our study. The solid green line corresponds to `moderate to strong' evidence on the Jeffrey's Scale and the solid red line corresponds to `decisive' evidence favouring flat \lam CDM. The PNGB and bimetric gravity linear interaction models are closest being decisively disfavoured whereas the growing $\nu$ mass, vacuum metamorphosis and quadratic bimetric are the least disfavoured.}
\end{figure}


The log Bayesian evidence decreases logarithmically with the increase in the prior volume (equation~\ref{eq:evidence}). Hence, having an arbitrarily large prior can, in extreme cases, lead to very low log evidence values. For example, an Einstein-de Sitter universe with \om\, = 1 would have similar evidence to flat \lam CDM with a U(0., 10$^{122}$) prior despite having an extremely poor fit to the data (although in this case, much of the prior region can be ruled out since it predicts no big bang). This is an interesting effect in cases where the prior range is not constrained by theory. 

For the parameter estimation in Section~\ref{sec:param_est}, we use a sampling efficiency of 0.8. For consistency, we use the same input values for evidence calculation. We note that the recommended sampling efficiency for evidence calculation is lower  \cite[typically $\sim$ 0.3][]{2009MNRAS.398.1601F} to have larger sampling ellipses. Lowering the sampling efficiency does not significantly change the evidence values for most cases, but in cases where it does, the \dlz\, of the model relative to \lam CDM increases, hence, indicating an even stronger preference for standard cosmology.

\begin{table*}
\caption{Interpreting the difference in log evidence using the ``Jeffrey's scale" as presented in \citep{2008ConPh..49...71T}. }
\begin{tabular}{|l|c|}
\hline\hline
\dlz\, & Strength of Evidence \\
\hline
\textless 1 & Inconclusive \\
1.0  - 2.5 & Weak/Moderate \\
2.5  - 5.0 & Moderate/Strong \\
\textgreater 5.0 & Significant/Decisive \\
\hline
\end{tabular}
\label{tab:dlogZ}
\end{table*}

\begin{table*}
\begin{minipage}{120mm}
\centering
\caption{Comparison of the output evidence from Importance Nested Sampling \citep{2013arXiv1306.2144F}. The input data includes JLA SNe and the CMB/BAO ratio, $f$. The errors on ln\,$Z$ in each case are \textless 0.1. }
\begin{tabular}{|l|l|c|c|c|c|c|}
\hline\hline
Model & Parameters & ln$Z_{f}$ & $\Delta$ & $\frac{Z_{m}}{Z_{f\Lambda}}$ & $\chi^2_{min}$ & Evidence Meaning\\
\hline
(flat) $\Lambda$CDM & \om & -359.6 & $\cdots$ & 1.000 & 685.7 & $\cdots$ \\ 
$\Lambda$CDM & \om, \ok &  -362.2 & -2.6  & 0.074 & 684.9  & Moderate/Strong \\ 
&&&&&&\\
Vacuum Metamorphosis\footnote{No assumption of flatness for this model, compare to \lam CDM without assumption of flatness, hence despite a higher \dlz\, than 1, the evidence would be inconclusive} & \om, \ost   & -361.1 & -1.5 & 0.223 & 683.0 
 &   Inconclusive/Weak \\
Doomsday & \om, \wo  & -362.7 & -3.1 & 0.045 & 684.8 & Moderate/Strong \\
Slow-Roll One parameter & \om, \dw\, & -361.4 & -1.8 & 0.150 & 684.9 & Weak/Moderate \\
PNGB  & \ophi, \wo, $K$ & -363.3 & -3.7 & 0.024 &  682.9  & Moderate/Strong  \\
Algebraic Thawing & \om, \wo, $p$  & -362.8 & -3.2 & 0.040 & 684.7 & Moderate/Strong \\
Algebraic Thawing (p=1) & \om, \wo & -362.8 & -3.2 & 0.040 & 684.7 & Moderate/Strong \\

Growing $\nu$ mass & \ome, \onu  &  -360.5 &  -0.9 & 0.405 & 684.5 & Inconclusive \\
Bimetric - Linear & \om &  -363.6 & -4.0 & 0.024 & 691.8 &  Moderate/Strong  \\
Bimetric - Quadratic & \om, $B_2$ & -360.1 & -0.5 & 0.606 & 685.6 & Inconclusive \\
&&&&&& \\

\hline
\end{tabular}
\label{tab:model_selec}
\end{minipage}
\end{table*}

\subsection{Model Comparison with CMB and BAO}
As noted in Section~\ref{sec:param_est}, some of the models in this study can also be constrained independently with CMB and BAO observations, detailed in Section~\ref{ssec:cmb} and \ref{ssec:bao}. The models for which the CMB compressed likelihood is suitable are algebraic thawing, PNGB, doomsday and one-parameter slow roll dark energy, all belonging to the class of thawing scalar field models. In this section we calculate the Bayes Factor for these models relative to the flat \lam CDM scenario (this is a justified comparison since all these models assume flatness). Since the CMB and BAO data combined in such a way is more precise than the CMB/BAO ratio, we use this investigation to see whether there are any strong preferences for the models tested here.
The resulting $\mathrm{ln}\, Z$ values are presented in Table~\ref{tab:model_selec_thaw}. 

\begin{table*} 
\begin{minipage}{120mm}
\centering
\caption{Evidence comparison for thawing models using CMB compressed likelihood (Section~\ref{ssec:cmb}) and BAO (Section~\ref{ssec:bao}). The errors on ln\,$Z$ in each case are \textless 0.1.}
\begin{tabular}{|l|l|c|c|c|c|c|}
\hline\hline
Model & Parameters &  $\mathrm{ln} Z_{f}$  & $\Delta$ & $\frac{Z_{f\Lambda}}{Z_{m}}$ & $\chi^2_{min}$ & Evidence Meaning\\
\hline
flat $\Lambda$CDM & \om  &  -362.6 & $\cdots$  & 1.00 & 686.7 & $\cdots$ \\ 
$\Lambda$CDM & \om  &  -365.3 & -2.7 & 0.067 & 686.3 & Moderate/Strong \\ 
&&&&&&\\

Doomsday & \om, \wo  & -368.7 & -6.1  &  0.002 & 686.3 & Significant/Decisive \\
Algebraic Thawing & \om, \wo, $p$  & -369.1 & -6.5 & 0.001 & 686.3 &   Significant/Decisive \\
Algebraic Thawing ($p =1$) & \om, \wo  & -369.1 & -6.5 & 0.001 & 686.4 & Significant/Decisive  \\
PNGB  & \ophi, \wo, $K$  & -368.5 & -5.9 & 0.003 & 686.1 &  Significant/Decisive \\
Slow-Roll One parameter & \om, \dw\, & -368.2 & -5.6 & 0.004 & 686.2  & Significant/Decisive \\
&&&&&& \\

\hline
\end{tabular}
\label{tab:model_selec_thaw}
\end{minipage}
\end{table*} 
We note that non-flat \lam CDM has moderate evidence against it, relative to the flat \lam CDM model. However, all thawing models have a \dlz\, of \textgreater 5 
which, on the Jeffrey's scale indicates decisive evidence in favour of standard \lam CDM. This is substantial against thawing behaviour of dark energy. 

Although there is a significant discussion about the applicability of compressed CMB likelihood \cite[for e.g., see][]{2016A&A...594A..14P}, we note that the results do not change significantly for only a combination of SNe~Ia and BAO, i.e. even in the absence of CMB data, there is a strong evidence against thawing models. For e.g. for the algebraic thawing model, we still obtain a Bayes Factor of $\sim$ 20 with only SN~Ia and BAO data.

\section{Forecasts for future surveys}
\label{sec:forecast}
In this section we look into the plausibility of future surveys to distinguish exotic cosmologies from standard \lam CDM. We take an example cosmology as the algebraic thawing model and compute the Bayes Factor for different values of \wo\, to determine for what input cosmologies can \lam CDM be ruled out. For our input datasets we use Type Ia supernovae forecast for the DESIRE survey which aims to extend the supernova Hubble diagram to z $\sim$ 1.5 \cite{2014A&A...572A..80A}. We also use complementary BAO and weak lensing (WL) from the Large Synoptic Survey Telescope \cite{2008arXiv0805.2366I}, BAOs from  DESI \cite{2016arXiv161100036D} and HETDEX \cite{2014JCAP...05..023F} as well as \hz\, measurements from WFIRST \cite{2012arXiv1208.4012G}, Euclid \cite{2010arXiv1001.0061R}, DESI \cite{2016arXiv161100036D} and HETDEX \cite{2014JCAP...05..023F}. Although the list of future surveys for the different cosmological probes used here is not exhaustive the sample of surveys is representative of the constraints expected from future surveys. 

\begin{equation}
\chi^2_{\mathrm{DESIRE}} = R^T W R,
\label{eq:chi_desire}
\end{equation}
where $R = d_M^{mod} - d_M^{obs}$ and $d_M = d_L  H_0/c$ \cite[see][for details of the expression]{2014A&A...572A..80A} and $W$ is piecewise weight matrix, detailed in Appendix E of \cite{2014A&A...572A..80A}. 
For the comoving and angular diameter distances from BAO and WL probes, the $\chi^2$ is simply given as 

\begin{equation}
\chi^2_{\mathrm{BAO/WL}} = (d_X^{\mathrm{mod}} - d_X^{\mathrm{obs}})^2/\sigma^2 ; X = A,C ,
\label{eq:bao_wl}
\end{equation}
where $A$ is the angular diameter distance and $C$ is the comoving distance. Similarly for the \hz\,  measurements

\begin{equation}
\chi^2_{H(z)} = (H^{\mathrm{mod}} - H^{\mathrm{obs}})^2/\sigma^2 ,
\label{eq:hz}
\end{equation}
describes the likelihood for obtaining the final constraints.

We aim to find for what parameter values of \wo\, can we exclude \lam CDM. Our exclusion criterion is a \dlz\, \textgreater 5, which corresponds to a Bayes Factor of $\sim$ 150, a strong exclusion criterion
From Table~\ref{tab:dlnz} we find that an algebraic thawing cosmology with \wo\, = -0.92 can be distinguished significantly from \lam CDM. 


\begin{table}
\centering
\caption{\dlz\, values for different baseline parameters of algebraic thawing (using all data combined). Values for \om\, and $p$  are 0.3 and 1.1 respectively}
\begin{tabular}{|l|c|}
\hline\hline
\wo & \dlz\\
\hline
-0.90 & 12.5 \\
-0.91 & 8.9 \\
-0.92 & 7.1\\
-0.93 & 4.4 \\
-0.94 & 2.2  \\
\hline\hline
\end{tabular}
\label{tab:dlnz}
\end{table}

In the above analysis we looked at how future data will be able to distinguish standard cosmology from more exotic models. 
Here, we also calculate the precision with which the present day EoS for the algebraic thawing model, assuming that it is the underlying cosmological model, can be measured by future cosmological probes.  The results are summarised in Table~\ref{tab:w0_forecast}.
The weakest constraints for a single low-redshift probes in combination with a CMB prior \cite{2016A&A...594A..14P} is from \hz\,  measurements. The BAOs perform slightly better than the SN constraints for the above mentioned combination of datasets. For a combination of all data, \wo\, can be constrained to within 2$\%$ which is a significant improvement on the current value and slightly better than the forecasts in  \cite{2015PhRvD..91f3006L}, due to the extensive BAO and \hz\, data used here. Similarly to  \cite{2015PhRvD..91f3006L} we find that low-redshift probes are critical to tightening the limits on \wo.

\begin{table}
\centering
\caption{68$\%$ confidence level precision estimates for \wo\,, the present day equation of state in the algebraic thawing model. See text for details about each probe.}
\begin{tabular}{|l|c|}
\hline\hline
Dataset & $\sigma(w_0)$\\
\hline
SN+CMB & 0.032 \\
BAO+CMB & 0.027 \\
$H(z)$ + CMB & 0.050 \\
SN+BAO+$H(z)$+CMB & 0.020 \\
\hline\hline
\end{tabular}
\label{tab:w0_forecast}
\end{table}

\begin{figure}
\includegraphics[width=.8\textwidth]{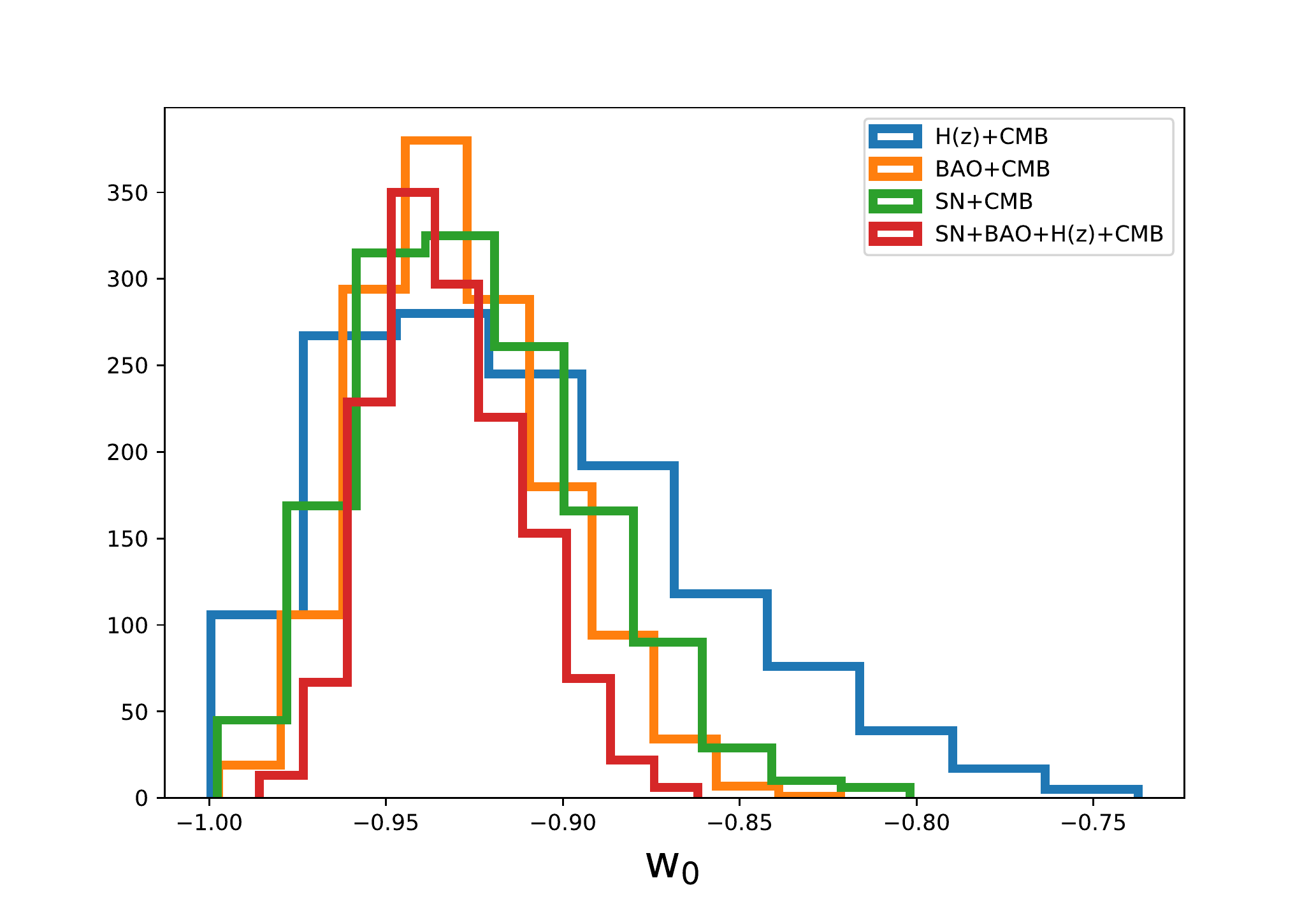}
\caption{Posterior distribution for future constraints on \wo\, from a combination of different cosmological probes. For the optimal case of all cosmological probes combined, \wo\, can be constrained to 2$\%$.}
\label{fig:forecast}
\end{figure}



\section{Discussion and Conclusions}
\label{sec:disc}

In this paper we have used the recent catalog of SN~Ia, estimate of the CMB first acoustic peak and the BAO scale to obtain constraints on exotic cosmological models beyond the standard \lam CDM scenario. We employed a model selection criterion (namely, the Bayesian evidence) to compare the different models. 

We found that most models fit the data well with a very similar goodness of fit criterion for each model. Models like the vacuum metamorphosis show possible signs of a deviation from standard behaviour, though the uncertainties in the measurements of \om\, and \ost\, are large and hence, the significance is only at the 1.5 $\sigma$ level. We also note \cite[as previously seen in ][]{2009ApJ...695..391R} that in the growing neutrino mass model there is a significant degeneracy between the early dark energy density. \ome,  and present day neutrino density \onu. Analysing this model with complete CMB, SN~Ia and BAO data will present interesting constraints on the present-day neutrino mass. 

We applied a model selection criterion, namely the logarithm of the Bayes Factor, to determine whether the data significantly prefer any of the models tested here. Using the CMB/BAO ratio combined with SN~Ia, we find that there is moderate to strong evidence suggesting that some of the thawing scalar field models, namely, algebraic thawing, PNGB, doomsday can be excluded (Table~\ref{tab:model_selec}). We note however that this is not an exhaustive list of models displaying thawing behaviour and other models like a quadratic or quartic potential can also be compared similarly \cite{2015PhRvD..91f3006L}. The bimetric gravity model with only the linear interaction term has the lowest evidence and is also a poor fit to the data. This model fits the SN~Ia and the CMB/BAO ratio independently, however, the two results are inconsistent, hence, the combined fit strongly disfavours this model. This illustrates the power of combining different geometric probes to gain information that no individual probe can provide. We note that this is the first study to present moderate to strong evidence for disfavouring thawing quintessence relative to concordance cosmology, although other extensions to standard cosmology, like time-varying dark energy, were shown to be marginally disfavoured in \cite{2006PhRvD..74l3506L,2007MNRAS.379..169S,2012JCAP...09..020V, 2016arXiv160306563S}. 

We note that some of the thawing models, like the PNGB model are motivated by technical naturalness to solve the fine-tuning problem, making them possibly even more well motivated than \lam. In this study, for a uniform comparison of the models tested, we used flat priors on $K$, the curvature of the potential at maximum. A possible extension of this analysis could involve using numerical calculations of the prior region on $K$ \cite[see][]{2017JCAP...01..023S} since the authors note a significant prior dependence for their results of a PNGB model. We aim to follow-up a comparison of such models with a complete CMB likelihood and test such a dependence in the future. 
Some models in our sample, namely, the vacuum metamorphosis and growing neutrino mass for which we cannot use the CMB compressed likelihood and hence only present parameter estimates and model comparison with the CMB/BAO ratio. Using the CMB and BAO probes individually can improve the parameter constraints presented here. 

When using the CMB compressed likelihood and BAO angular scale as separate probes, we find that the thawing scalar field models can be strongly excluded (Table~\ref{tab:model_selec_thaw}). The other models in our study cannot be tested with the compressed likelihood since they are either motivated by modified gravity or include an early dark energy term \cite{2016A&A...594A..14P}. 
We also tested extensions of the \lam\,CDM that include curvature as a free parameter. As in previous studies \cite{2016arXiv160703155A,2014A&A...564A.125B} we find the data to be consistent with a flat universe (\ok\, = 0.). However, we find that this model is moderately disfavoured relative to the flat \lam CDM case. This result is consistent with previous studies that use different algorithms to calculate the evidence \cite{2009MNRAS.398.1601F,2009MNRAS.397..431V,2010MNRAS.405.2381K}. 
We find that the SNe~Ia magnitude-redshift relation and the CMB/BAO ratio contribute nearly equally to distinguish the different models. 


Although the current data only moderately distinguish between standard cosmology and more exotic extensions, we find that future data can decisively rule out certain models.  We forecast the errors on the parameters for one class of thawing models (algebraic) to determine how well can standard \lam CDM can be distinguished from algebraic thawing cosmology. We find that a strong distinction can be provided between thawing models with \wo\, = -0.92 and standard \lam CDM with future SN,BAO, CMB and H(z) data. For this case, the present day EoS can be constrained with an error $\sigma (w_0)$ = 0.02, which presents a promising prospect to constrain any dynamical behaviour of dark energy.

{\it Acknowledgements:} 
We would like to thank Zachary Slepian, Marc Betoule, Robert Caldwell for fruitful discussions and Eric Linder for help at the start of the project and comments on a previous draft. We would like to thank Daniel Mortlock for discussions on Bayesian Evidence, Johannes Buchner and Vlas Sokolov for inputs on nested sampling. This work was funded through grants from the Knut and Alice Wallenberg foundation, the Swedish Research Council (VR) and the Swedish National Space Board.

\bibliographystyle{unsrt}
\bibliography{biblio,SNIa}

\begin{thebibliography}{100}
\bibitem{2009PhRvD..80l3001K}
M.~{Kunz}.
\newblock {Degeneracy between the dark components resulting from the fact that
  gravity only measures the total energy-momentum tensor}.
\newblock {\em \prd}, 80(12):123001, December 2009.
\end{thebibliography}

\end{document}